\documentclass[floats,floatfix,showpacs,amssymb,prd,twocolumn,superscriptaddress,nofootinbib]{revtex4-1}

\usepackage{amssymb,amsmath,verbatim,mathtools,needspace,enumitem,etoolbox,graphicx,physics,microtype,afterpage,tabularx,color,textcmds}

\usepackage[dvipsnames, usenames]{xcolor}

\definecolor{linkcolor}{rgb}{0.0,0.3,0.5}
\usepackage[unicode, colorlinks=true, linkcolor=linkcolor, citecolor=linkcolor, filecolor=linkcolor,urlcolor=linkcolor, pdfusetitle]{hyperref}
\usepackage[all]{hypcap}
\usepackage[T1]{fontenc}
\usepackage[utf8]{inputenc}
\usepackage{aas_macros}
\newcommand\Wider[2][3em]{%
\makebox[\linewidth][c]{%
  \begin{minipage}{\dimexpr\textwidth+#1\relax}
  \raggedright#2
  \end{minipage}%
  }%
}

\makeatletter

\@addtoreset{equation}{section}
\makeatother

\begin{document}

\title{Post-Newtonian dynamics and black hole thermodynamics\\ in Einstein-scalar-Gauss-Bonnet gravity}
\author{F\'elix-Louis Juli\'e}
\email{fjulie@jhu.edu}
\affiliation{Department of Physics and Astronomy, Johns Hopkins University, 3400 N. Charles Street, Baltimore, MD 21218, USA}

\author{Emanuele Berti}
\email{berti@jhu.edu}
\affiliation{Department of Physics and Astronomy, Johns Hopkins University, 3400 N. Charles Street, Baltimore, MD 21218, USA}

\date{\today}

\begin{abstract}
We study the post-Newtonian dynamics of black hole binaries in Einstein-scalar-Gauss-Bonnet gravity theories. To this aim we build static, spherically symmetric black hole solutions at fourth order in the Gauss-Bonnet coupling $\alpha$. We then ``skeletonize'' these solutions by reducing them to point particles with scalar field-dependent masses, showing that this procedure amounts to fixing the Wald entropy of the black holes during their slow inspiral. The cosmological value of the scalar field plays a crucial role in the dynamics of the binary. We compute the two-body Lagrangian at first post-Newtonian order and show that no regularization procedure is needed to obtain the Gauss-Bonnet contributions to the fields, which are finite.  We illustrate the power of our approach by Pad\'e-resumming the so-called ``sensitivities,'' which measure the coupling of the skeletonized body to the scalar field, for some specific theories of interest.
\end{abstract}

\maketitle

\section{Introduction}

The quest for a quantum theory of gravity and observational puzzles in modern cosmology have led to several proposals for theories of gravity that differ from general relativity (GR). By Lovelock's theorem, these modifications of GR almost inevitably lead to additional degrees of freedom, and the simplest and best studied extensions involve scalar fields (see e.g.~\cite{Berti:2015itd}).

The recent LIGO/Virgo observations of gravitational waves finally allow us to test the presence of these additional degrees of freedom and their effect in the strong-field gravity regime. Binary black holes (BHs) have several advantages as probes of strong-field gravity. First of all, observations of binary BH mergers outnumber those of other compact binaries involving neutron stars, at least so far. Furthermore, BHs allow us to perform ``cleaner'' tests of gravity than systems involving matter, because we do not need to make assumptions on the poorly known state of matter at supranuclear densities.

Unfortunately, the simplicity of BHs in GR applies also to the structure and dynamics of BHs in modified theories of gravity: stringent no-hair theorems imply that BH mergers in many of these theories are observationally indistinguishable from GR (see~\cite{Herdeiro:2015waa} for a review of no-hair theorems). For example, one such no-hair theorem implies that static, asymptotically flat BH solution are the same as in GR for a vast majority of scalar-tensor theories that lead to second-order equations of motion~\cite{Hui:2012qt}.

This no-hair theorem is violated in Einstein-scalar-Gauss-Bonnet (EsGB) gravity, a theory where a scalar degree of freedom $\varphi$ couples to the Gauss-Bonnet scalar $\mathcal R^2_{\rm GB}=R^{\mu\nu\rho\sigma}R_{\mu\nu\rho\sigma}-4R^{\mu\nu}R_{\mu\nu}+R^2$. EsGB gravity is exceptional in many ways: a coupling of the form $f(\varphi)\mathcal R^2_{\rm GB}$ allows for nontrivial effects in the strong-field, large curvature regime, even in four-dimensional spacetimes.

In fact, the existence of hairy BH solutions in such theories has been known for a long time. Early studies focused on Einstein-dilaton-Gauss-Bonnet (EdGB) gravity~\cite{Kanti:1995vq,Pani:2009wy}, the low-energy effective action of the bosonic sector of heterotic string theory~\cite{Gross:1986mw}. More recently, BH solutions have been found for more general coupling functions~\cite{Antoniou:2017acq}.
Even the simplest (shift-symmetric) Gauss-Bonnet theories~\cite{Sotiriou:2013qea,Sotiriou:2014pfa,Maselli:2015yva}, where $f(\varphi)\propto \varphi$, were shown to evade the no-hair theorems of~\cite{Hui:2012qt}.

A no-hair theorem for stationary, asymptotically flat BHs in scalar-Gauss-Bonnet theories for a massless scalar with no self-interactions holds under the following conditions: the function $f(\varphi)$ must have an extremum at some constant $\varphi=\bar\varphi$, i.e. $f'(\bar\varphi)=0$, and
$f''(\bar\varphi)\mathcal{G}<0$.
When only the latter condition is violated -- e.g., when $f(\varphi)\propto \varphi^2$~\cite{Silva:2017uqg} -- these theories exhibit \emph{spontaneous BH scalarization}, i.e. they allow for nontrivial scalar field configurations that reduce to the BHs of GR in the appropriate limit~\cite{Doneva:2017bvd,Silva:2017uqg}.
The stability of these solutions was studied in various recent works~\cite{Blazquez-Salcedo:2018jnn,Minamitsuji:2018xde,Silva:2018qhn,Macedo:2019sem}.

Whenever BHs are endowed with scalar ``hair'', BH binaries produce dipolar radiation in the early inspiral, and their merger dynamics is also different from GR~\cite{Horbatsch:2011ye,Yagi:2011xp,Healy:2011ef,Stein:2013wza,Berti:2013gfa,Barausse:2016eii,Prabhu:2018aun,Berti:2018cxi,Witek:2018dmd}.
These considerations led to analytical and numerical work on the dynamics of BH binaries in EsGB gravity at lowest order in the coupling~\cite{Yagi:2011xp,Witek:2018dmd}.
Ref.~\cite{Yagi:2011xp} computed the dipolar energy flux treating the conservative dynamics at leading (Newtonian) order and, therefore, assuming that the scalar charges of the binary component are constant.

We improve on that treatment in two ways: (1) we allow for the fact that the BH masses and scalar ``charges'' are not constant: instead, we consistently skeletonize the BHs following a well-established procedure first introduced by Eardley in scalar-tensor gravity~\cite{1975ApJ...196L..59E}, and recently generalized to Einstein-Maxwell-dilaton theory by one of us~\cite{Julie:2017rpw,Julie:2018lfp}; (2) as a consequence of the skeletonization, we can self-consistently compute higher-order post-Newtonian (PN) terms in the Lagrangian.

The plan of the paper is as follows.  In Sec.~\ref{sectionBHetThermo} we find analytical solutions for hairy black holes valid up to fourth order in the GB coupling, and we discuss their thermodynamical properties.
In Sec.~\ref{sectionSkeletonization} we use Eardley's ``skeletonization'' technique to show that the mass is not constant, and therefore that it is necessary to go beyond Newtonian order in the conservative dynamics. We also find the remarkable result that, in the PN regime, a BH can be uniquely characterized by its Wald entropy.
In Sec.~\ref{sectionBlackHoleBinaries} we present the two-body Lagrangian for a generic EsGB theory of gravity and, as an example, we discuss BH sensitivities in EdGB.
In Sec.~\ref{sec:conclusions} we conclude by pointing out possible directions for future work.

Some lengthy technical material is relegated to the appendices.
Appendix~\ref{appFieldEqn} presents a simple derivation of the EsGB field equations in arbitrary dimensions that (as far as we know) does not appear in the published literature.
Appendix~\ref{appendixFullBHsolution} lists some of the lengthier coefficients in the analytical expansion of the metric and scalar field for EsGB BHs at fourth order in the GB coupling.
Appendix~\ref{appThermoParameters} gives analytical expressions for the thermodynamical variables characterizing these BHs.
Appendix~\ref{AppendixPNLagrangian}  contains the derivation of one of our most important results: the two-body Lagrangian at first post-Newtonian (1PN) order. Along the way, we find another remarkable result: the Gauss-Bonnet contributions to the fields are {\em finite}, and no regularization procedure is necessary at 1PN order.
In Appendix~\ref{sec:AppQuadraticShiftSymmetric} we study the BH sensitivities in two special cases of EsGB gravity that were extensively considered in the literature: theories where the coupling depends quadratically on the field and shift-symmetric theories.

\section{Hairy black holes and thermodynamics\label{sectionBHetThermo}}
EsGB theories supplement GR with a massless scalar field coupled to the Gauss-Bonnet Lagrangian density. In vacuum and in geometrical units ($G\equiv c\equiv 1$), they are described by the action
\begin{equation}
  I=
  \int\!\frac{d^4x\sqrt{-g}}{16\pi}\bigg(R-2g^{\mu\nu}\partial_\mu\varphi\partial_\nu\varphi+\alpha f(\varphi)\mathcal R^2_{\rm GB}\bigg)\ ,\label{vacuumAction}
\end{equation}
where $R$ is the Ricci scalar, $g=\det g_{\mu\nu}$ denotes the metric determinant, and the integral of the Gauss-Bonnet scalar over spacetime $\int\! d^Dx\sqrt{-g}\mathcal R^2_{\rm GB}$ is a boundary term in dimension $D\leqslant 4$ (see e.g.~\cite{Myers:1987yn, Deruelle:2018vtt}). The coupling constant $\alpha$ (which is chosen to be positive without loss of generality) has dimensions of length squared, and $f(\varphi)$ is a dimensionless function defining the theory.

The vacuum field equations follow from the variation of the action (\ref{vacuumAction}):
\begin{subequations}
\begin{align}
R_{\mu\nu}\!&=\!2\partial_\mu\varphi\partial_\nu\varphi-4\alpha\!\left(\!P_{\mu\alpha\nu\beta}-\frac{g_{\mu\nu}}{2}P_{\alpha\beta}\!\right)\!\nabla^\alpha\nabla^\beta\! f,\label{einsteinVacFieldEqn}\\
\Box\varphi&=-\frac{1}{4}\alpha f'(\varphi)\mathcal R_{\rm GB}^2\ ,\label{KleinGordonVacFieldEqn}
\end{align}
\label{vacFieldEqn}%
\end{subequations}
where $\nabla_\mu$ denotes the covariant derivative associated to $g_{\mu\nu}$, and $\Box\equiv\nabla^\mu\nabla_\mu$. The divergenceless quantity $P_{\mu\nu\rho\sigma}=R_{\mu\nu\rho\sigma}-2g_{\mu[\rho}R_{\sigma]\nu}+2g_{\nu[\rho}R_{\sigma]\mu}+g_{\mu[\rho}g_{\sigma]\nu}R$ has the symmetries of the Riemann tensor (see e.g. \cite{Davis:2002gn, Deruelle:2003ps}), and $P_{\mu\nu}\equiv P^\lambda_{\ \,\mu\lambda\nu}$. Details of the derivation of Eq.~(\ref{einsteinVacFieldEqn}) are in Appendix \ref{appFieldEqn} (see also \cite{Kanti:1995vq} and \cite{Ripley:2019irj} for alternative formulations of the EsGB field equations).

\subsection{Black holes in generic Einstein-scalar-Gauss-Bonnet theories\label{sectionTNsolution}}

There is an extensive body of work on BHs in EsGB gravity. When the coupling $\alpha$ between the scalar field and the Gauss-Bonnet invariant is small, the vacuum field equations (\ref{vacFieldEqn}) can be solved analytically and perturbatively around GR. This program was carried out in the string-inspired EdGB theory with coupling $f(\varphi)=\frac{1}{4} e^{2\varphi}$ to find static solutions~\cite{Mignemi:1992nt,Torii:1996yi,Yunes:2011we} and their slowly spinning counterparts~\cite{Ayzenberg:2014aka,Pani:2011gy,Maselli:2015tta} up to order $\mathcal O(\alpha^7)$. The same approximation scheme was used in the ``shift-symmetric'' theory $f(\varphi)=2\varphi$ (which is invariant under $\varphi\rightarrow\varphi+\rm{constant}$, see (\ref{vacuumAction}) and below), but only for nonspinning BHs and up to order $\mathcal O(\alpha^2)$~\cite{Sotiriou:2013qea,Sotiriou:2014pfa}.

The field equations (\ref{vacFieldEqn}) were solved numerically and nonperturbatively, also for rapidly spinning BHs (see e.g. \cite{Kanti:1995vq,Pani:2009wy,Kleihaus:2015aje}). Theories where $f'(\varphi)=0$ and $f''(\varphi) {\mathcal R}_{\rm GB}^2>0$ for some $\varphi=\varphi_0$ -- such as the theories $f(\varphi)=\frac{\varphi^2}{2}(1+\lambda \varphi^2)$ and $f(\varphi)=-\frac{1}{2\lambda}e^{-\lambda\varphi^2}$, with $\lambda\in I\!\!R$ -- predict instabilities of GR BHs in favor of other branches of stable solutions with nontrivial scalar ``hair''~\cite{Silva:2017uqg,Doneva:2017bvd,Antoniou:2017acq,Silva:2018qhn,Minamitsuji:2018xde,Macedo:2019sem,Minamitsuji:2019iwp,Cunha:2019dwb}.

Our first goal is to complement and extend these results by obtaining analytical, asymptotically flat BH solutions with (secondary) scalar ``hair'' at high order in the coupling $\alpha$ and \textit{in an arbitrary EsGB theory}
(see e.g.~\cite{Herdeiro:2015waa} for a review of no-hair theorems and the classification of hairy BH solutions).
Let us focus on static, spherically symmetric solutions in a Just coordinate system:
\begin{equation}
ds^2=-A(r)\,dt^2+\frac{dr^2}{A(r)}+ B(r)\,r^2(d\theta^2+\sin^2\!\theta\, d\phi^2) ,\label{SSSansatz}
\end{equation}
with $\varphi=\varphi(r)$.
For a Schwarzschild spacetime with mass parameter $m$ we have $A=1-2m/r$, $B=1$ and $\varphi=\varphi_\infty$, where $\varphi_\infty$ is an arbitrary constant.

When the coupling constant $\alpha$ is nonzero, we must solve Eq.~(\ref{vacFieldEqn}) as a perturbative expansion in the dimensionless parameter
\begin{equation}
\epsilon\equiv \frac{\alpha f'(\varphi_\infty)}{4m^2}\ll 1\ ,
\end{equation}
such that $\varphi-\varphi_\infty=\mathcal O(\epsilon)$: cf. Eq.~(\ref{KleinGordonVacFieldEqn}).
The leading-order EsGB correction to GR is straightforward.
The right-hand side of the Einstein equations (\ref{einsteinVacFieldEqn}) vanishes at order $\mathcal O(\epsilon)$, so the Schwarzschild metric is still the solution, which sources the scalar field through the Kretschmann scalar: $\mathcal R^2_{\rm GB}=48m^2/r^6+\mathcal O(\epsilon^2)$.
At higher orders the calculation proceeds as follows. Substitute the ansatz (\ref{SSSansatz}), with
\begin{subequations}
\begin{align}
  A&=1-\frac{2m}{r}+\sum_{i=1}^{4}\epsilon^i A_i(r)+\mathcal O(\epsilon^5),\\
  B&=1+\sum_{i=1}^{4}\epsilon^i B_i(r)+\mathcal O(\epsilon^5),\\
  \varphi&=\varphi_\infty+\sum_{i=1}^{4}\epsilon^i \varphi_i(r)+\mathcal O(\epsilon^5),
\end{align}
\end{subequations}
together with the Taylor expansion
\begin{equation}
  f(\varphi)=\sum_{n=0}^{4}\frac{1}{n!}f^{(n)}(\varphi_\infty)(\varphi-\varphi_\infty)^n+\mathcal O(\epsilon^5),
\end{equation}
in the field equations (\ref{vacFieldEqn}) and solve order-by-order, ignoring branches with singular horizons.
The result is:
\begin{subequations}
\begin{align}
  A&=1-u-\epsilon^2\left(\frac{u^3}{3}-\frac{11u^4}{6}+\frac{u^5}{30}+\frac{17u^7}{15}\right)\nonumber\\
  &+\epsilon^3 A_3+\epsilon^4 A_4+\mathcal O(\epsilon^5)\ ,\label{BHsolutionA}\\
  B&=1-\epsilon^2\left(u^2+\frac{2u^3}{3}+\frac{7u^4}{6}+\frac{4u^5}{5}+\frac{3u^6}{5}\right)\nonumber\\
  &+\epsilon^3 B_3+\epsilon^4 B_4+\mathcal O(\epsilon^5)\ ,\label{BHsolutionB}\\
  \varphi &=\varphi_\infty+\epsilon \left(u+\frac{u^2}{2}+\frac{u^3}{3}\right)\nonumber\\
  &+\epsilon^2 \varphi_2+\epsilon^3 \varphi_3+\epsilon^4 \varphi_4+\mathcal O(\epsilon^4)\ ,\label{BHsolutionPhi}
\end{align}\label{BHsolution}%
\end{subequations}
with $u\equiv 2m/r$. For convenience, the EsGB corrections to the coefficients in $A$
which are proportional to $1/r$ have been conveniently reabsorbed in the definition of $m$.
The quantities $A_{i\geqslant 3}$, $B_{i\geqslant3}$ and $\varphi_{i\geqslant 2}$ depend on $m$ and on the function $f(\varphi)$ and its derivatives evaluated at infinity, i.e. $(d^nf/d\varphi^n)(\varphi_\infty)$ with $n\in [0,\,4]$.  They are rather lengthy, and their explicit expressions can be found in Eqs.~(\ref{BHsolutionA})--(\ref{BHsolutionPhiApp}) of Appendix~\ref{appendixFullBHsolution}.

The solution above depends on two integration constants: the Arnowitt-Deser-Misner (ADM) mass $m$ -- i.e., one-half the $\mathcal O(1/r)$ coefficient of $g_{rr}$ at infinity -- and the asymptotic value $\varphi_\infty$ of the scalar field at spatial infinity.

The results above match previous analytical work in the respective limits, but they also extend it in several ways:

\begin{itemize}
\item[(i)] the solution (\ref{SSSansatz}) with the expansion coefficients listed in Eqs.~(\ref{BHsolution}) is valid for arbitrary EsGB coupling functions;
\item[(ii)] the solution is given explicitly at order $\mathcal O(\epsilon^4)$ in the Gauss-Bonnet coupling, and in principle it can be extended to higher orders. As such it contains detailed information on the BH's structure, that will be useful below to characterize the dynamics of BH binaries (cf. Sec.~\ref{PNparameters}).
\item[(iii)] the solution depends on the asymptotic value $\varphi_\infty$ of the scalar field at infinity, unlike most previous work on isolated BHs, where $\varphi_\infty$ was (and could be) set to zero for simplicity: see e.g. \cite{Yunes:2011we,Silva:2018qhn}. For binary BHs, \textit{$\varphi_\infty$ cannot be fixed to zero} anymore. This is one of the key messages of this paper, for reasons explained in Sec.~\ref{sectionSkeletonization} below.
\end{itemize}

\subsection{Black hole thermodynamics\label{sectionThermo}}

The solution given in Eqs.~(\ref{SSSansatz}) and (\ref{BHsolution})
can be checked to have the properties expected of a BH spacetime.  First of all, in Eq.~(\ref{eq:Kretschmann}) of Appendix~\ref{appendixFullBHsolution} we show that the Kretschmann curvature invariant is finite everywhere outside the horizon, where the horizon radius $r_{\rm H}$ is trivially defined as the outermost zero of $A(r)$ in the Just coordinates of Eq.~(\ref{SSSansatz}).

Perhaps more remarkably, the EsGB BH solution satisfies the first law of BH thermodynamics in terms of the following intensive and extensive parameters.

The BH temperature $T$ is
\begin{equation}
T\equiv \frac{\kappa}{2\pi}\ ,\label{temperature}
\end{equation}
where the surface gravity $\kappa$ is defined by $\kappa^2\equiv -\frac{1}{2}(\nabla_\mu\xi_\nu\nabla^\mu\xi^\nu)_{r_{\rm H}}$, and $\xi^\mu=(1,0,0,0)$ is the timelike Killing vector associated to stationarity.

The action (\ref{vacuumAction})
can be written in terms of a Lagrangian density $\mathcal L$
as $I\equiv\int\! d^4x\sqrt{-g}\mathcal L$.
The BH entropy $S_{\rm w}$ is then given by Wald's formula \cite{Wald:1993nt}:
\begin{align}
S_{\rm w}&\equiv-8\pi\int_{r_{\rm H}}\!\!\!\! d\theta d\phi\sqrt{\sigma}\frac{\partial\mathcal L}{\partial R_{\mu\nu\rho\sigma}}\epsilon_{\mu\nu}\epsilon_{\rho\sigma}\ .\label{waldEntropy1}
\end{align}
Here $\sigma$ is the determinant of the induced metric on the horizon with unit normal vectors $n^\mu=(1/\sqrt{-g_{tt}},0,0,0)$ and $l^\mu=(0,1/\sqrt{g_{rr}},0,0)$, and $\epsilon_{\mu\nu}=n_{[\mu}l_{\nu]}$. Evaluating the Wald entropy (\ref{waldEntropy1}) for the action (\ref{vacuumAction}) yields
\begin{align}
S_{\rm w}=\frac{\mathcal A_{\rm H}}{4} + 4\alpha\pi f(\varphi_{\rm H})\ ,\label{waldEntropy2}
\end{align}
i.e. the total entropy is the sum of the standard Bekenstein entropy $S_{\rm B}=\mathcal A_{\rm H}/4$ and a Gauss-Bonnet contribution~\cite{Maeda:2009uy}.
Here $\varphi_{\rm H}\equiv\varphi(r_{\rm H})$ denotes the value of the scalar field on the horizon.
 
Finally, it is well-known in scalar-tensor theories that the scalar field contributes to the global mass $M$ as follows:
\begin{equation}
M=m+\int\! D\, d\varphi_\infty\ ,\label{katzMass}
\end{equation}
where $m$ is the ADM mass defined earlier; see e.g. \cite{Gibbons:1996af,Cardenas:2016uzx,Anabalon:2016ece,Cardenas:2017chu} and references therein. The quantity $D$ is defined from an asymptotic expansion of the scalar field as $\varphi=\varphi_\infty+D/r+\mathcal O(1/r^2)$, and it is sometimes called the scalar ``charge'' of the BH, although $\varphi$ is not a gauge field in general.

We can now evaluate the temperature $T$, entropy $S_{\rm w}$ and ``charge'' $D$ for our analytical BH solution. Their expressions in terms of the integration constants $m$ and $\varphi_\infty$ are collected in %
Appendix~\ref{appThermoParameters},
and they can be used to check that the variation of $S_{\rm w}$ and $M$ with respect to both $m$ and $\varphi_\infty$ satisfy the following identity, at least at order $\mathcal O(\epsilon^4)$:
\begin{equation}
T\delta S_{\rm w}=\delta M\ .\label{firstLaw}
\end{equation}
This first law of BH thermodynamics describes how the equilibrium configuration of the EsGB BH readjusts when it interacts with its environment. In particular, in Sec.~\ref{sectionBlackHoleBinaries} below we will investigate the variations of the scalar field environment $\varphi_\infty$ induced by a far-away binary companion.

To summarize: we have solved the vacuum field equations (\ref{vacFieldEqn}), obtained a BH solution at fourth order in the coupling $\alpha$, and verified that this solution satisfies a first law of BH thermodynamics that accounts for the scalar field environment $\varphi_\infty$ of the BH, when the BH entropy is defined \`a la Wald. These results are our starting point for an analytical investigation of the dynamics of BH binaries in a generic EsGB theory.

\section{Skeletonization: reducing an Einstein-scalar-Gauss-Bonnet black hole to a point particle\label{sectionSkeletonization}}

We now want to describe the motion of EsGB BHs in binary systems. To this aim, it is convenient to ``skeletonize'' the BH by adding it as a point source $A$ to the vacuum action (\ref{vacuumAction}):
\begin{align}
&I_{\rm pp}[g_{\mu\nu},\varphi,x_A^\mu]=I-\int\! m_A(\varphi)\, ds_A\ .\label{skelAction}
\end{align}
Here $ds_A=\sqrt{-g_{\mu\nu}dx_A^\mu dx_A^\nu}$, and $x_A^\mu[s_A]$ is the worldline of particle A. The mass function $m_A(\varphi)$, which replaces the constant GR ``mass'' $m_A$, is a scalar function that depends on the value of the scalar field at its location $x_A^\mu(s_A)$, and it was first introduced by Eardley to account for the coupling of a star $A$ to its scalar field environment in scalar-tensor theories~\cite{1975ApJ...196L..59E}. This approach was generalized to ``hairy'' BHs in Einstein-Maxwell-dilaton (EMD) theories in~\cite{Julie:2017rpw,Julie:2018lfp} (see also \cite{Khalil:2018aaj}).

The ansatz (\ref{skelAction}) does not depend on any field gradients, e.g. $\partial_\mu\varphi$. Neglecting such terms corresponds to neglecting finite-size effects (e.g., tidal forces)~\cite{Damour:1998jk}: cf.~\cite{Bernard:2019yfz} for recent work on this topic in scalar-tensor theories.

The question we address here is the calculation of the function $m_A(\varphi)$ for EsGB BHs. Following the techniques developed in \cite{Julie:2017rpw}, we impose that the fields generated by extremizing the action~(\ref{skelAction}) match those of the BH built in the previous section.

\subsection{The matching conditions}

The field equations following from the variation of (\ref{skelAction}) are:
\begin{subequations}
\begin{align}
  R_{\mu\nu}&=2\partial_\mu\varphi\partial_\nu\varphi-4\alpha\left(P_{\mu\alpha\nu\beta}-\frac{1}{2}g_{\mu\nu}P_{\alpha\beta}\right)\nabla^\alpha\nabla^\beta f(\varphi)\nonumber\\
              &+8\pi\left(T^A_{\mu\nu}-\frac{1}{2}g_{\mu\nu}T^A\right)\ ,\label{einsteinFieldEqnSkel}\\
  \Box\varphi&=-\frac{1}{4}\alpha f'(\varphi)\mathcal R_{\rm GB}^2\nonumber\\
  &+4\pi\!\int\! ds_A\frac{dm_A}{d\varphi}\frac{\delta^{(4)}(x-x_A(s_A))}{\sqrt{-g}}\ ,
               \label{KleinGordonFieldEqnSkel}
\end{align}
\label{fieldEqnSkel}%
\end{subequations}
where $\delta^{(4)}(x-y)$ is the four-dimensional Dirac distribution and $T_{\mu\nu}^A$ is the distributional stress-energy tensor
\begin{equation}
T_A^{\mu\nu}=\int\! ds_A\, m_A(\varphi)\frac{\delta^{(4)}(x-x_A(s_A))}{\sqrt{-g}}\frac{dx_A^\mu}{ds_A}\frac{dx_A^\nu}{ds_A}\ .\label{TmunuA}
\end{equation}

Let us solve the filed equations perturbatively around a Minkowski metric $\eta_{\mu\nu}$ and a constant scalar background $\varphi_\infty$. At infinity and at leading order, the Gauss-Bonnet contributions to the right-hand side of Eq.~(\ref{fieldEqnSkel}) vanish. In the rest frame of the point-source $A$ (i.e. setting $\mathbf{x}_A=\mathbf{0}$) and using harmonic coordinates such that $\partial_\mu(\sqrt{-\tilde g}\tilde g^{\mu\nu})=0$ we find:
\begin{subequations}
\begin{align}
\tilde g_{\mu\nu}&=\eta_{\mu\nu}+\delta_{\mu\nu}\left(\frac{2m_A(\varphi_\infty)}{\tilde r}\right)+\mathcal O\left(\frac{1}{\tilde r^2}\right)\ ,\\
\varphi &=\varphi_\infty-\frac{1}{\tilde r}\frac{dm_A}{d\varphi}(\varphi_\infty)+\mathcal O\left(\frac{1}{\tilde r^2}\right)\ .
\end{align}\label{asymptSkel}%
\end{subequations}
At leading order, the large-${\tilde r}$ expansion of the metric and of the scalar field depends on the function $m_A(\varphi_\infty)$, its derivative $m'_A(\varphi_\infty)$, and the asymptotic scalar field value $\varphi_\infty$.
This should be compared with the asymptotic behavior of the BH spacetime we derived in Sec.~\ref{sectionTNsolution} written in terms of the same harmonic radial coordinate ${\tilde r}$ through the relation $r=\tilde r+m+\mathcal O(1/\tilde r)$:
\begin{subequations}
\begin{align}
\tilde g_{\mu\nu}&=\eta_{\mu\nu}+\delta_{\mu\nu}\left(\frac{2m}{\tilde r}\right)+\mathcal O\left(\frac{1}{\tilde r^2}\right)\ ,\\
\varphi &=\varphi_\infty+\frac{D}{\tilde r}+\mathcal O\left(\frac{1}{\tilde r^2}\right)\ .
\end{align}\label{asymptBH}%
\end{subequations}

Therefore the skeletonized point particle $A$ will match the fields of an EsGB BH if and only if
\begin{subequations}
\begin{align}
m_A(\varphi_\infty)&=m\ ,\label{mathcingM}\\
m'_A(\varphi_\infty)&=-D\ .\label{mathcingMPrime}
\end{align}\label{mathcing}%
\end{subequations}
Indeed, when seen as a boundary condition at infinity, Eqs.~(\ref{asymptBH}) identify a unique solution to the vacuum field equations (\ref{vacFieldEqn}). Therefore, a point particle with a scalar field-dependent mass $m_A(\varphi)$ satisfying the matching conditions (\ref{mathcing}) generates fields which reproduce (outside of the distribution) those of the BH {\em at all orders} in a $1/r$ expansion. The covariance of Eq.~(\ref{skelAction}) ensures that this is true in any reference frame, that is, independently of the motion of the BH.\\

Now, since the scalar hair of EsGB BHs is secondary (see e.g.~\cite{Herdeiro:2015waa}) $D$ is not an independent integration constant, and it can be written as a function $D(m,\varphi_\infty)$: cf. Eq.~(\ref{scalarChargeMPHI}). 
We can replace $m$ by $m_A(\varphi_\infty)$ in $D(m,\varphi_\infty)$ because of the matching condition (\ref{mathcingM}), and replace the resulting expression on the right-hand side of the matching condition (\ref{mathcingMPrime}). This procedure yields the following differential equation for the function $m_A(\varphi)$:
\begin{widetext}
\begin{align}
\frac{m'_A(\varphi)}{m_A(\varphi)}&+2\epsilon_A(\varphi)+\epsilon_A(\varphi)^2\frac{73  f''(\varphi)}{30f'(\varphi)
   }+\epsilon_A(\varphi)^3 \left(\frac{73 }{15}+\frac{12511 
   f''(\varphi)^2}{3780f'(\varphi)^2}+\frac{12511  f'''(\varphi)}{7560f'(\varphi)}\right)\label{eqDiffSkel}\\
 &+\epsilon_A(\varphi)^4\left(\frac{227192473 
  f''(\varphi)^3}{49896000 f'(\varphi)^3}+\frac{31557593  f''(\varphi) f'''(\varphi)}{4536000f'(\varphi)^2}+\frac{143467 f''(\varphi)}{4158f'(\varphi)}+\frac{799607
   f''''(\varphi)}{997920  f'(\varphi)}\right)+\cdots=0\ ,\nonumber
\end{align}
\end{widetext}
where $\epsilon_A(\varphi)\equiv \alpha f'(\varphi)/(4m_A(\varphi)^2)$, and where we dropped the $\infty$ subscript for simplicity.  We now turn to the solution of this first-order differential equation, which will involve a single integration constant $\mu_A$. As we show below, this constant is related to the Wald entropy of the BH.

\subsection{A constant-entropy skeletonization}
\label{sec:WaldEntropy}

The solution to (\ref{eqDiffSkel}) can be built iteratively and reads
\begin{align}
  m_A(\varphi)&=\mu_A  \left(1-\frac{\alpha f(\varphi)}{2\mu_A^2}-\frac{\alpha^2 F_2(\varphi)}{\mu_A^4}\right. \nonumber\\
                &-\left. \frac{\alpha^3  F_3(\varphi)}{\mu_A^6}-\frac{\alpha^4  F_4(\varphi)}{\mu_A^8}+\cdots\right)\ , \label{sensitivity}
\end{align}
where $\mu_A$ is a positive integration constant with dimensions of mass. The theory-dependent functions $F_i(\varphi)$ depend on $f(\varphi)$ and its derivatives:
\begin{subequations}
\begin{align}
 &F_2(\varphi)=\frac{f(\varphi)^2}{8}+\frac{73 f'(\varphi)^2}{960}\ , \\
& F_3(\varphi)=\frac{f(\varphi)^3}{16}+\frac{73 f(\varphi) f'(\varphi)^2}{640}+\frac{12511 f'(\varphi)^2 f''(\varphi
   )}{483840 }\ ,\\
   &  F_4(\varphi)=\frac{5 f(\varphi)^4}{128}+\frac{73
   f(\varphi)^2 f'(\varphi)^2}{512}+\frac{12534857 f'(\varphi)^4}{425779200} \nonumber \\
  &+\frac{12511 f(\varphi) f'(\varphi)^2 f''(\varphi)}{193536} +\frac{227192473 f'(\varphi)^2 f''(\varphi)^2}{25546752000}\nonumber \\
  &+\frac{799607 f'(\varphi)^3 f'''(\varphi)}{255467520}\ .
\end{align}\label{sensitivityF}%
\end{subequations}

The mass function $m_A(\varphi)$ of an EsGB BH, Eq.~(\ref{sensitivity}), is the main result of this section. The information encoded in the complicated form of the spacetime metric -- cf. Eqs.~(\ref{BHsolution}), (\ref{BHsolutionA}) and (\ref{BHsolutionPhi}) -- is now summarized in a set of four compact body-independent functions $F_i(\varphi)$, which will turn out to play an important role in describing the interaction of the BH with a companion. More importantly, the expression of $m_A(\varphi)$ shows that a skeletonized BH is characterized by a single parameter $\mu_A$.

The physical interpretation of this parameter can be found thus: invert Eq.~(\ref{sensitivity}) order-by-order in $\alpha$, and use the matching condition (\ref{mathcingM}) to write $\mu_A$ as a function of $m$ and $\varphi_\infty$. The result shows that $\mu_A$ is nothing but the BH's irreducible mass~\cite{Christodoulou:1970wf}:
 \begin{equation}
\mu_A=M_{\rm irr}=\sqrt{\frac{S_{\rm w}}{4\pi}}\ ,\label{irreducibleMass}
\end{equation}
where $S_{\rm w}$ is the BH's Wald entropy defined earlier, and computed explicitly in Appendix~\ref{appThermoParameters} [cf. Eq.~(\ref{entropyMPHI})].

The reason why the BH's (Wald) entropy plays such a central role in the skeletonization is best revealed by thermodynamics. The variation of the global mass $M$ defined in (\ref{katzMass}),
\begin{equation}
\delta M=\delta m+D\,\delta\varphi_\infty\ ,\label{thermoVSskel}
\end{equation}
vanishes identically because of the matching conditions (\ref{mathcing}): $\delta M=0$. In other words, when we skeletonize a BH representing it by a point particle $A$, we implicitly assume that it is isolated, i.e., that it exchanges no mass $M$ with its environment. By the first law (\ref{firstLaw}), the BH entropy must then remain constant: $\delta S_{\rm w}=0$. Therefore it is a suitable parameter to characterize the BH.

The physical meaning of the ``skeletonization'' process can be interpreted as follows. When replaced by a point particle, a BH is described by a constant (Wald) entropy $S_{\rm w}$ together with a scalar environment $\varphi_\infty$ which {\em cannot be set to zero}: for example, in Sec.~\ref{sectionBlackHoleBinaries} the value of $\varphi_\infty$ will be determined by a (far-away) companion $B$. During the bodies' slow inspiral, the variation of $\varphi_\infty$ forces BH $A$ to readjust its equilibrium configuration \textit{adiabatically}, i.e. at constant values of its Wald entropy $S_{\rm w}$. On the contrary, the BH's ADM mass $m$ and scalar ``charge'' $D$ are from now on functions of $\varphi_\infty$: cf. Eq.~(\ref{mathcing}).

Previous work~\cite{Julie:2017rpw,Cardenas:2017chu} applied a similar skeletonization procedure to Einstein-Maxwell-dilaton (EMD) BHs, characterized by a scalar ``hair'' along with a $U(1)$ charge. The EMD mass function $m_A(\varphi)$ was also found to depend on a single integration constant (the irreducible mass $\mu_A^{\rm EMD}=\sqrt{S_{\rm B}/4\pi}$); however in the EMD case, $S_{\rm B}=\frac{1}{4}\mathcal A_{\rm H}$ is the Bekenstein entropy. This paper hints at a possible universality of this result, since its holds for theories whose metric sector differs from the Einstein-Hilbert action, as long as the Bekenstein entropy is replaced by Wald's. We conjecture that this conclusion might apply to any scalar-tensor theory of gravity.

\section{Black hole binaries\label{sectionBlackHoleBinaries}}

So far we found analytic solutions for isolated EsGB BHs, and skeletonized the BHs by describing them as point particles endowed with a scalar field-dependent mass $m_A(\varphi)$ which encodes information on their structure.
We can now describe a binary BH system by an action depending on two such mass functions $m_A(\varphi)$ ($A=1,\,2$):
\begin{align}
&I_{\rm pp}[g_{\mu\nu},\varphi,\{x_A^\mu\}]=I-\sum_A\int\! m_A(\varphi)\, ds_A\ ,\label{skelAction2body}
\end{align}
where we recall that $ds_A=\sqrt{-g_{\mu\nu}dx_A^\mu dx_A^\nu}$.

Starting from the skeleton action above, in Sec.~\ref{sectionPNlagr} we present the PN two-body Lagrangian for arbitrary compact binaries in EsGB theories (relegating the details of the calculation to Appendix~\ref{AppendixPNLagrangian}).  In Sec.~\ref{PNparameters} we use the mass function $m_A(\varphi)$ of Eq.~(\ref{sensitivity}) to better understand the dynamics of binaries composed of two ``hairy'' BHs in a specific class of EsGB theories: EdGB gravity. 

\subsection{The post-Newtonian Lagrangian\label{sectionPNlagr}}

The variation of (\ref{skelAction2body}) yields the field equations
\begin{subequations}
\label{fieldEqnSkel2}
\begin{align}
  R_{\mu\nu}&=2\partial_\mu\varphi\partial_\nu\varphi-4\alpha\left(P_{\mu\alpha\nu\beta}-\frac{1}{2}g_{\mu\nu}P_{\alpha\beta}\right)\nabla^\alpha\nabla^\beta f(\varphi)\nonumber\\
  &+8\pi\sum_A\left(T^A_{\mu\nu}-\frac{1}{2}g_{\mu\nu}T^A\right)\ ,\label{einsteinFieldEqnSkel}\\
  \Box\varphi&=-\frac{1}{4}\alpha f'(\varphi)\mathcal R_{\rm GB}^2\noindent\nonumber\\
  &+4\pi\sum_A\int\! ds_A\frac{dm_A}{d\varphi}\frac{\delta^{(4)}(x-x_A(s_A))}{\sqrt{-g}}\ ,\label{KleinGordonFieldEqnSkel}
\end{align}
\end{subequations}
where $T^A_{\mu\nu}$ denotes the distributional stress-energy tensor of particle $A$: cf. Eq.~(\ref{TmunuA}).

In this paper we focus on the conservative dynamics of compact binaries on bound orbits. When the bodies' relative orbital velocity $v$ is small and in the weak-field limit $m/r\ll 1$ (where $r$ is the orbital separation radius and $m$ their mass), the motion can be described in the PN framework.
In Appendix \ref{AppendixPNLagrangian} we derive the first PN two-body Lagrangian up to order $\mathcal O(v^2)\sim\mathcal O(m/r)$ beyond Newton. We solve the field equations  (\ref{fieldEqnSkel2}) perturbatively around a flat Minkowski metric $\eta_{\mu\nu}$ with a constant background scalar field value $\varphi_0$. As we shall illustrate below, $\varphi_0$ cannot be set to zero: its value is imposed by the binary's cosmological environment.

Adopting the conventions of Damour and Esposito-Far\`ese \cite{Damour:1992we,Damour:1995kt}, the mass functions $m_A(\varphi)$ and $m_B(\varphi)$ can be expanded by defining
\begin{align}
  \alpha_A(\varphi)&\equiv\frac{d\ln m_A(\varphi)}{d\varphi}\ ,
    \label{DefAlpha} \\
  \beta_A(\varphi)&\equiv\frac{d\alpha_A(\varphi)}{d\varphi}\ ,
    \label{DefBeta}
\end{align}
so that
\begin{align}
  m_A(\varphi)&= m_A^0 \left[1+\alpha_A^0 (\varphi-\varphi_0)\right. \label{eq:mAexp}\\
              &+\left.\frac{1}{2}({\alpha_A^0}^2+\beta_A^0)(\varphi-\varphi_0)^2+\mathcal O(\varphi-\varphi_0)^3\right]\ ,\nonumber
\end{align}
where from now on a ``$0$'' superscript means that the corresponding quantity is evaluated at $\varphi=\varphi_0$. The ``sensitivity'' $\alpha_A^0=(m_A'/m_A)(\varphi_0)$ measures the (relative) coupling of the skeletonized body $A$ to the scalar field -- see e.g. Eq.~(\ref{mathcing}) -- and it will play a key role below.

With these definitions, and working in a harmonic coordinate system such that $\partial_\mu(\sqrt{-g}g^{\mu\nu})=0$, the PN Lagrangian reads (reinstating Newton's constant $G$ for clarity):
\begin{widetext}
\begin{align}
L_{AB}&=-m_A^0-m_B^0+\frac{1}{2}m_A^0 \mathbf{v}_A^2+\frac{1}{2}m_B^0 \mathbf{v}_B^2+\frac{G_{AB}m_A^0m_B^0}{r}+\frac{1}{8}m_A^0\mathbf{v}_A^4+\frac{1}{8}m_B^0\mathbf{v}_B^4 \label{1PNLagrangian}\\
&+\frac{G_{AB}m_A^0m_B^0}{r}\left[\frac{3}{2}(\mathbf{v}_A^2+\mathbf{v}_B^2)-\frac{7}{2}(\mathbf v_A\cdot \mathbf v_B)-\frac{1}{2}(\mathbf n\cdot \mathbf v_A)(\mathbf n\cdot \mathbf v_B)+\bar\gamma_{AB}(\mathbf v_A-\mathbf v_B)^2\right]\nonumber\\
&-\frac{G_{AB}^2m_A^0m_B^0}{2r^2}\left[m_A^0(1+2\bar\beta_B)+m_B^0(1+2\bar\beta_A)\right]+\Delta L_{AB}^{\rm GB}+\mathcal O(v^6)\ ,\nonumber
\end{align}
\end{widetext}
where the Gauss-Bonnet contribution reads
\begin{align}
  \Delta L_{AB}^{\rm GB}&=\frac{\alpha f'(\varphi_0)}{r^2} \frac{G^2m_A^0m_B^0}{r^2}\nonumber\\
                          &\times \left[m_A^0(\alpha_B^0+2\alpha_A^0)+m_B^0(\alpha_A^0+2\alpha_B^0)\right].\label{deltaL-GB}
\end{align}
Here $\mathbf x_A$ denotes the position of body $A$, $r\equiv|\mathbf x_A-\mathbf x_B|$, $\mathbf n\equiv (\mathbf x_A-\mathbf x_B)/r$, and $\mathbf v_A\equiv d\mathbf x_A/dt$. We also introduced the combinations
\begin{subequations}
\begin{align}
  G_{AB}&\equiv G(1+\alpha_A^0\alpha_B^0)\ , \\
  \bar\gamma_{AB}&\equiv -2\frac{\alpha_A^0\alpha_B^0}{1+\alpha_A^0\alpha_B^0}\ , \\
  \bar\beta_A&\equiv\frac{1}{2}\frac{\beta_A^0{\alpha_B^0}^2}{(1+\alpha_A^0\alpha_B^0)^2}\ ,
\end{align}\label{defCombinaisonsPN}%
\end{subequations}
together with their counterparts that can be obtained by swapping indices ($A\leftrightarrow B$).

The two-body Lagrangian (\ref{1PNLagrangian}), including the Gauss-Bonnet contribution (\ref{deltaL-GB}), is one of the main results of this paper. It describes the first relativistic corrections to the dynamics of an arbitrary binary system in EsGB theories. The simplicity of the result is quite striking: $L_{AB}$ is the sum of the ordinary scalar-tensor two-body Lagrangian (see e.g.~\cite{Damour:1992we}) plus a term resulting from the complex coupling to the Gauss-Bonnet scalar, which (as shown in detail in Appendix~\ref{AppendixPNLagrangian}) boils down to adding the simple correction of Eq.~(\ref{deltaL-GB}).

Since $\Delta L^{\rm GB}_{AB}$ depends on an extra dimensionful coupling $\alpha$, it should \textit{a priori} be considered as a 1PN contribution to the two-body Lagrangian. However, by rewriting (\ref{deltaL-GB}) as
\begin{align}
  \Delta L_{AB}^{\rm GB}&=\frac{\alpha f'(\varphi_0)}{(GM^0)^2}\left(\frac{GM^0}{r}\right)^2\frac{G^2m_A^0m_B^0}{r^2}\nonumber\\
  &\times \left[m_A^0(\alpha_B^0+2\alpha_A^0)+m_B^0(\alpha_A^0+2\alpha_B^0)\right]\ ,\label{DeltaLAB}
\end{align}
with $M^0=m_A^0+m_B^0$, we can regard it as a 3PN correction whenever the ``small-$\alpha$'' approximation $\alpha f'(\varphi_0)\lesssim (GM^0)^2$ holds. This perturbative approximation is commonly used in the literature, and it was used in the derivation of our BH solutions (Sec.~\ref{sectionTNsolution}).

The two-body Lagrangian was recently calculated at 3PN order for pure scalar-tensor theories in \cite{Bernard:2018hta,Bernard:2018ivi}. Our results extend this Lagrangian to EsGB theories: we just need to add the contribution coming from Eq.~(\ref{DeltaLAB}). At least in the small-$\alpha$ regime, the results of Ref.~\cite{Bernard:2018ivi}, supplemented by the Gauss-Bonnet contribution (\ref{DeltaLAB}), yield the full EsGB Lagrangian at 3PN order.

In previous analytical calculations of the dynamics of binary systems in EsGB gravity~\cite{Yagi:2011xp} the field equations were sourced by particles with constant masses and constant scalar ``charges'', denoted by $m_A$ and $q_A$: see e.g. Eqs.~(63)-(64) or Eq.~(71) of~\cite{Yagi:2011xp}. This is equivalent to truncating the expansion (\ref{eq:mAexp}) at linear order. The work of~\cite{Yagi:2011xp} describes the conservative dynamics at leading (Newtonian) order. By endowing the particles with scalar-field dependent masses $m_A(\varphi)$, our treatment differs from theirs in two crucial ways:

\begin{itemize}
\item[(1)] we allow for the fact that the masses and scalar ``charges'' are not constant, as discussed below Eq.~(\ref{thermoVSskel});

\item[(2)] the skeletonization allows us to deal with higher PN terms: the  $\beta_A^0$-dependent contributions in Eq.~(\ref{1PNLagrangian}) cannot be captured by the approach of Ref.~\cite{Yagi:2011xp}.  \end{itemize}
 
The coupling $\alpha$ to the Gauss-Bonnet scalar affects the structure of the two-body Lagrangian only through the term (\ref{DeltaLAB}). However, $\alpha$ also crucially affects the masses $m_A(\varphi)$, and hence the values of the parameters $m_A^0$, $\alpha_A^0$, $\beta_A^0$
which appear also in the ``ordinary'' scalar-tensor part of the Lagrangian (\ref{1PNLagrangian}). In the next section we will study $\alpha_A(\varphi_0)$ for several selected EsGB coupling functions, using the corresponding BH solutions and their skeletonization (Sec.~\ref{sectionSkeletonization}).

\subsection{Black hole sensitivities in a binary system:\\the Einstein-dilaton-Gauss-Bonnet example\label{PNparameters}}

In subsection~\ref{sectionPNlagr} we derived a PN two-body Lagrangian which generalizes that of GR through the quantities 
$\alpha_A$ and $\beta_A$ defined in Eqs.~(\ref{DefAlpha}) and (\ref{DefBeta}).  We now specialize this Lagrangian to a binary of EsGB BHs.
More precisely, our goal is to compute the ``sensitivity parameter'' $\alpha_A^0=\alpha_A(\varphi_0)$ associated to BH $A$, which  is characterized by a constant irreducible mass $\mu_A$ (cf. Sec.~\ref{sec:WaldEntropy}). The quantities $\alpha^0_{A}$ play a central role: once we know $\alpha^0_{A}$ we can easily obtain $\beta_A^0$, and quadratic combinations of $\alpha_A^0$ [$G_{AB}$, $\bar\gamma_{AB}$ and $\bar\beta_{A}$ : cf. Eq.~(\ref{defCombinaisonsPN})] drive all EsGB corrections to GR.

Taking the logarithmic derivative of $m_A(\varphi)$ given in (\ref{sensitivity}) yields
\begin{equation}
\alpha_A^0=-\frac{x}{2}-x^2 A_2(\varphi_0)-x^3 A_3(\varphi_0)-x^4 A_4(\varphi_0)+\mathcal O(x^5)\ ,
\label{scalarChargeUnexpanded}
\end{equation}
where
\begin{equation}
  x\equiv\frac{\alpha f'(\varphi_0)}{\mu_A^2}
\end{equation}
is the weak GB coupling of a constant-entropy BH. The functions $A_i(\varphi_0)$ depend on the theory  only through the function $f(\varphi_0)$ and its derivatives $f^{(n)}(\varphi_0)$:
\begin{widetext}
\begin{subequations}
\begin{align}
A_2(\varphi_0)&=\frac{f(\varphi_0 )}{2 f'(\varphi_0 )}+\frac{73 f''(\varphi_0 )}{480 f'(\varphi_0 )}\ ,\\
A_3(\varphi_0)&=\frac{73}{480}+\frac{f(\varphi_0 )^2}{2 f'(\varphi_0 )^2}+\frac{73 f(\varphi_0 ) f''(\varphi_0 )}{240 f'(\varphi_0
   )^2}+\frac{12511 f''(\varphi_0 )^2}{241920 f'(\varphi_0 )^2}+\frac{12511 f^{(3)}(\varphi_0 )}{483840
   f'(\varphi_0 )}\ ,\\
A_4(\varphi_0)&=\frac{f(\varphi_0 )^3}{2 f'(\varphi_0 )^3}+\frac{73 f(\varphi_0 )^2 f''(\varphi_0 )}{160 f'(\varphi_0
   )^3}+\frac{5505779 f''(\varphi_0 )}{26611200 f'(\varphi_0 )}+\frac{227192473 f''(\varphi_0
   )^3}{12773376000 f'(\varphi_0 )^3}\\
   &+\frac{31557593 f''(\varphi_0 ) f^{(3)}(\varphi_0 )}{1161216000
   f'(\varphi_0 )^2}+\frac{73 f(\varphi_0 )}{160 f'(\varphi_0 )}+\frac{12511 f(\varphi_0 ) f''(\varphi_0
   )^2}{80640 f'(\varphi_0 )^3}+\frac{12511 f(\varphi_0 ) f^{(3)}(\varphi_0 )}{161280 f'(\varphi_0
   )^2}+\frac{799607 f^{(4)}(\varphi_0 )}{255467520 f'(\varphi_0 )}\ .\nonumber
\end{align}\label{scalarCouplingA}
\end{subequations}
\end{widetext}
Moreover, in the following it will be convenient to resum the Taylor expansion (\ref{scalarChargeUnexpanded}) in the variable $x$ by using a diagonal $(2,2)$ Pad\'e approximant (see e.g.~\cite{Damour:1997ub,2002nrca.book.....P} for discussions of Pad\'e approximants):
\begin{equation}
  \alpha_{A,\text{Pad\'e}}^0=\mathcal P^2_{\ 2}[\alpha_A^0, x]\ ,
  \label{padeResum}
\end{equation}
The Pad\'e resummation, which replaces polynomials by rational functions, has two important advantages: it can improve the convergence of the expansion (\ref{scalarChargeUnexpanded}) and (perhaps more importantly) it can capture interesting nonperturbative phenomena, as we clarify below. Using Eqs.~(\ref{scalarChargeUnexpanded}) and (\ref{padeResum}) we shall identify regimes where the BH binary dynamics significantly departs from GR.

In the remainder of this section we focus on EdGB gravity as a prototypical, well motivated special case of EsGB theories. To improve readability, we relegate two other important examples (quadratic and shift-symmetric theories) to Appendix~\ref{sec:AppQuadraticShiftSymmetric}.

\subsubsection*{Einstein-dilaton-Gauss-Bonnet theories\label{subsectionEdGB}}

Using the conventions of Ref.~\cite{Kanti:1995vq}, the ``string-inspired" subclass of EdGB theories is characterized by the exponential coupling function
\begin{equation}
f(\varphi)=\frac{e^{2\varphi}}{4}\ ,\label{fDilaton}
\end{equation}
so that the fundamental action (\ref{vacuumAction}) is invariant under the simultaneous redefinitions $\varphi\to\varphi+\Delta\varphi$ and $\alpha\to\alpha e^{-2\Delta\varphi}$, with $\Delta\varphi$ an arbitrary constant. Recall that here the parameter $\alpha$ (with no subscripts) denotes the fundamental coupling to the GB invariant in the action (\ref{vacuumAction}).

The scalar coupling function for BH $A$ of Eq.~(\ref{scalarChargeUnexpanded}) then becomes
\begin{equation}
\alpha_A^0=-\frac{x}{2}-\frac{133}{240 } x^2
   -\frac{35947 }{40320}x^3-\frac{474404471
   }{266112000}x^4+\mathcal O\left(x^5\right)\, ,
   \label{scalarCouplingDilaton}
\end{equation}
with
\begin{equation}
 x=\frac{\alpha e^{2\varphi_0}}{2\mu_A^2}\,.
\end{equation}
This sensitivity preserves the symmetry of the fundamental action, in the sense that it is symmetric under the transformation $\varphi_0\to\varphi_0+\Delta\varphi$, $\ \alpha\to\alpha e^{-2\Delta\varphi}$.

\begin{figure*}[th]
  \includegraphics[width=\columnwidth]{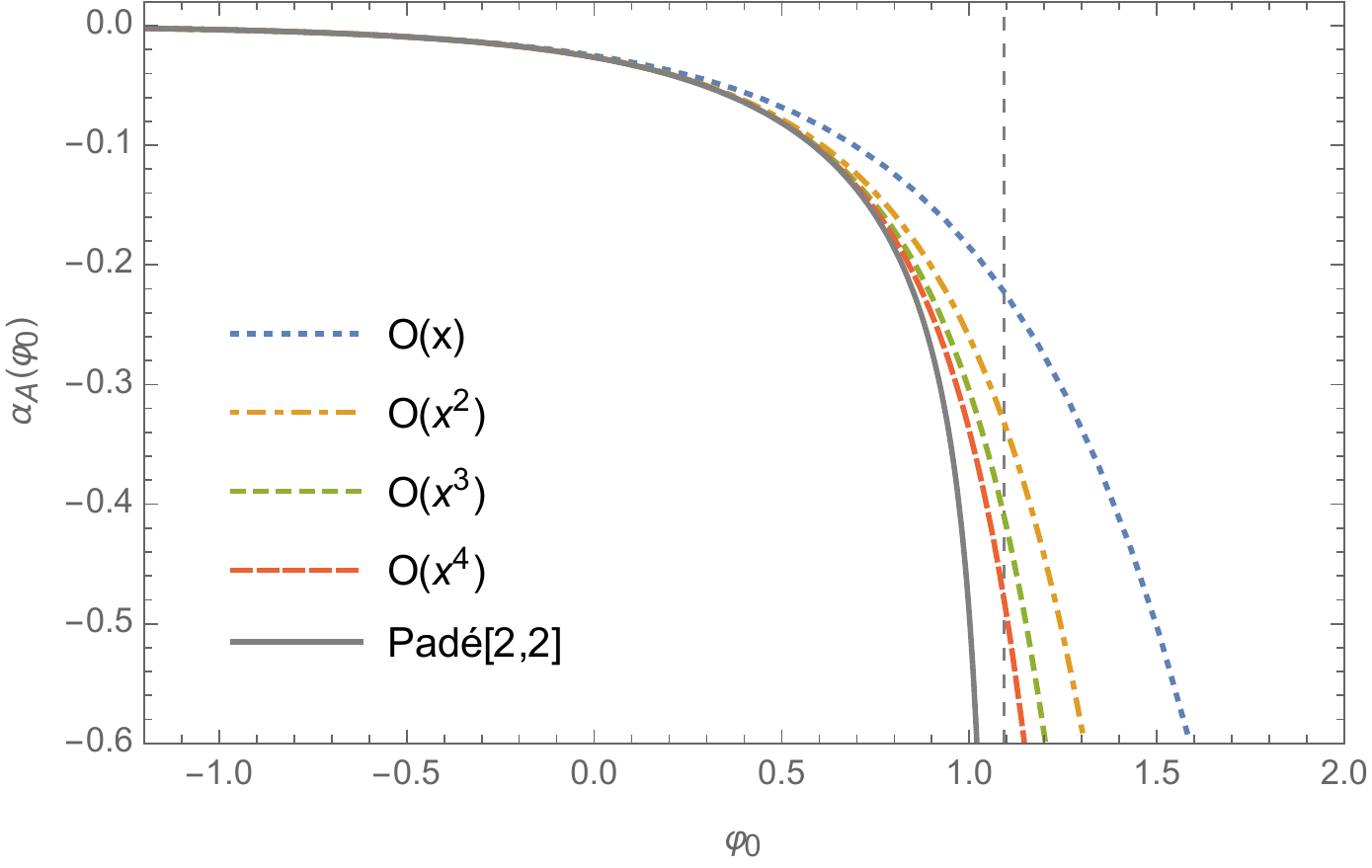}
  \includegraphics[width=\columnwidth]{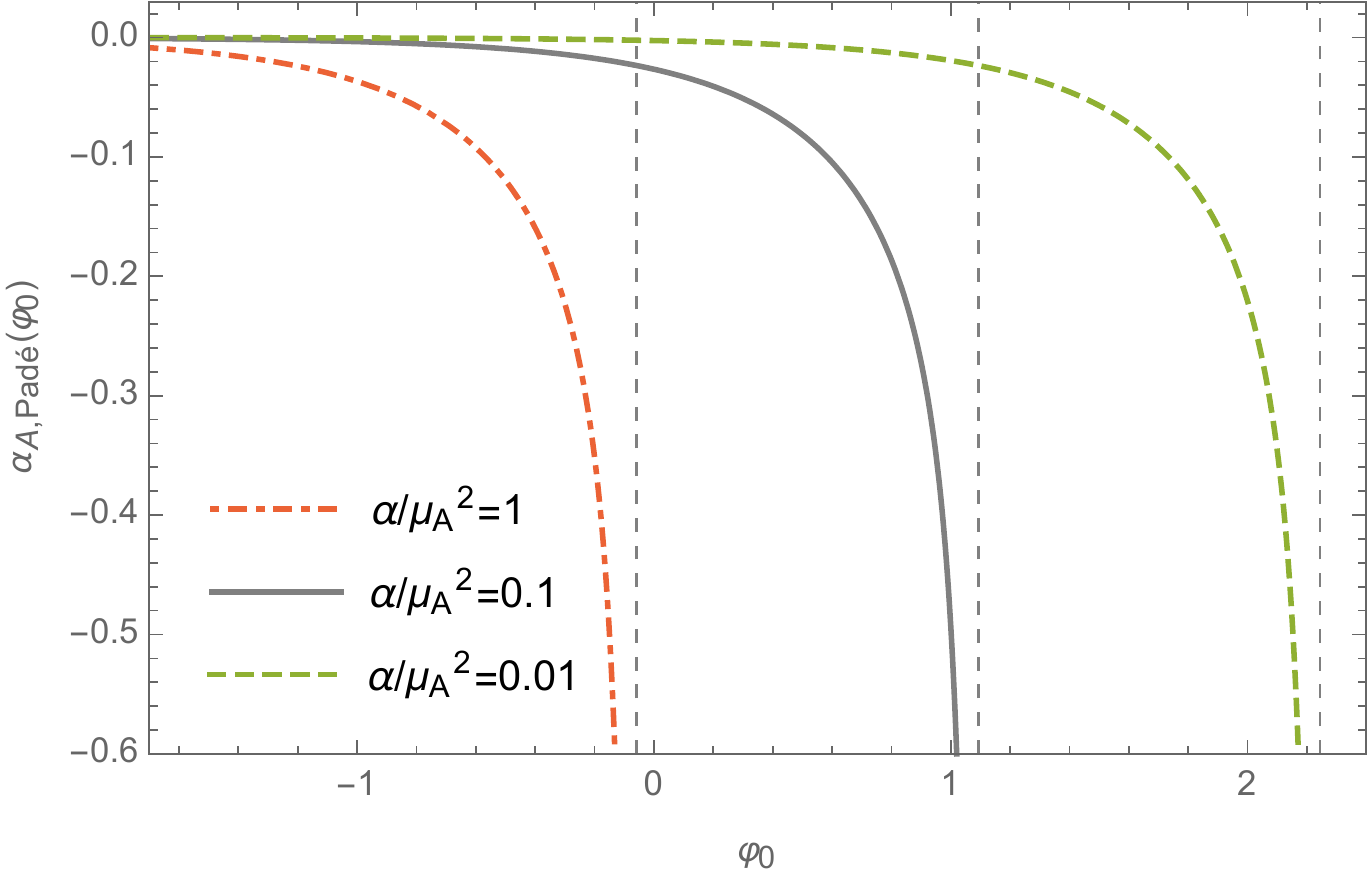}
  \caption{Sensitivity $\alpha_A^0=\alpha_A(\varphi_0)$ of EdGB BHs as a function of the cosmological scalar field $\varphi_0$. The left panel shows various truncations of the Taylor series (\ref{scalarCouplingDilaton}) for a BH with $\alpha/\mu_A^2=0.1$. The right panel shows the $(2,2)$ Pad\'e resummation $\alpha_{A,\text{Pad\'e}}^0$ of Eq.~(\ref{padeResum}) for three different BHs with $\alpha/\mu_A^2=\{1,0.1,0.01\}$. The Pad\'e resummation improves the convergence of $\alpha_A^0$, and it predicts the existence of a pole at $x_{\rm pole}=0.445$ (dashed vertical lines).}
\label{EdGB_scalarCoupling}
\end{figure*}

The left panel of Fig.~\ref{EdGB_scalarCoupling} shows various approximants of the series (\ref{scalarCouplingDilaton}) truncated at order $\mathcal O(x^n)$ as a function of $\varphi_0$, setting $\alpha/\mu_A^2=0.1$. The expansion coefficients in Eq.~(\ref{scalarCouplingDilaton}) are all negative, so the series diverges at large $\varphi_0$, with a slope which increases with the truncation order $n$.

To accelerate the convergence of our expansion, we Pad\'e-resum it as in Eq.~(\ref{padeResum}). This operation reveals a remarkable feature: the resummed sensitivity $\alpha_{A,\text{Pad\'e}}^0$, also shown in the left panel of Fig.~\ref{EdGB_scalarCoupling}, has a pole at
\begin{equation}
x_{\rm pole}=\frac{\alpha e^{2\varphi_0^{\rm pole}}}{2\mu_A^2}=0.445\ .\label{EdGBpoleValue}
\end{equation}

The presence of a pole in the full, nonperturbative coupling $\alpha_A^0$ is at first sight surprising. Upon further consideration, however, this feature is particularly appealing.
While no exact analytical BH solutions are known in EsGB theories, it is well-known that the area $\mathcal A_{\rm H}$ of a static BH and the value $\varphi_{\rm H}$ of the scalar field at the horizon must satisfy the following nonperturbative constraint (see e.g. \cite{Doneva:2017bvd}):
\begin{equation}
24\,\alpha^2 f'(\varphi_{\rm H})^2<\left(\frac{\mathcal A_{\rm H}}{4\pi}\right)^2 .\label{BHbound}
\end{equation}
When the constraint is violated, the scalar field diverges at the horizon and the BH becomes a ``naked singularity" (cf.~\cite{Doneva:2017bvd,Kanti:1995vq}, or~\cite{Kleihaus:2011tg} for further numerical evidence).
In the EdGB subclass of theories studied here, and for a skeletonized BH characterized by a constant irreducible mass $\mu_A$, we can use Eqs.~(\ref{waldEntropy2}) and (\ref{irreducibleMass}) to write $\mathcal A_{\rm H}$  in term of $\mu_A$ and $\varphi_{\rm H}$, so the constraint above becomes
\begin{equation}
\frac{\alpha e^{2\varphi_{\rm H}}}{2\mu_A^2}<\frac{2}{1+\sqrt{6}}\ .\label{BHboundEdGB}
\end{equation}
This nonperturbative bound confirms the conclusion that an EdGB BH solution with fixed Wald entropy $\mu_A$ must become singular when the scalar field at the horizon $\varphi_{\rm H}$ reaches a critical value.

Unfortunately, the prediction (\ref{EdGBpoleValue}) for the numerical value of the pole cannot be directly compared with the nonperturbative condition (\ref{BHboundEdGB}). Such a comparison would require us to relate the value $\varphi_{\rm H}$ of the scalar field on the horizon to the value $\varphi_0$ of the field at infinity.\footnote{An approximate relation between $\varphi_{\rm H}$ and $\varphi_0$ can be found from the solution (\ref{BHsolutionPhi}) for the scalar field and the horizon location (\ref{horizonLocation}), which are both known in the perturbative limit (i.e., for small coupling). Using Eq.~(\ref{sensitivity}) we can write $m_A$ in terms of $\mu_A$. Inserting the resulting $\varphi_{\rm H}(\varphi_0)$ in (\ref{BHboundEdGB}) then yields $x_{\rm pole}=0.331$. Considering that the results of Sec.~\ref{sectionTNsolution} break down in the nonperturbative regime, this value is at least in qualitative agreement with Eq.~(\ref{EdGBpoleValue}). As another indication of convergence, we checked that the diagonal, $(2,2)$ Pad\'e approximant performs ``better'' than off-diagonal Pad\'e approximants, in the sense that the pole location (\ref{EdGBpoleValue}) predicted by the diagonal approximant is the closest to the value $x_{\rm pole}=0.331$ that results from the procedure described here.} It is still significant that the Pad\'e resummation predicts the existence of a critical value for $\varphi_0$ at which the BH sensitivity $\alpha_A^0$ diverges.

Figure~\ref{EdGB_scalarCoupling} highlights the crucial role of the (cosmological) background scalar field $\varphi_0$ on the dynamics of an EdGB BH binary.
As $\varphi_0$ increases, the BH transitions progressively between two ``universal'' regimes:

\begin{itemize}
\item[(i)] a {\em decoupled} regime where the BH is indistinguishable from a Schwarzschild BH in GR, since both $\alpha_A^0$ and $\beta_A^0=d\alpha_A/d\varphi(\varphi_0)$ (as well as higher-order derivatives of $\alpha_A^0$) vanish; and

\item[(ii)] a regime with $\alpha_A^0\to -\infty$ (and $\beta_A^0\to -\infty$) where the BH is {\em strongly coupled} to the scalar field, inducing large deviations to the GR two-body Lagrangian through $\bar \gamma_{AB}$ and $\bar\beta_{A/B}$ [cf. Eqs.~(\ref{defCombinaisonsPN})].
\end{itemize}

This ``transition'' is universal because the Wald entropy $\mu_A$ only affects the location of the pole, as shown in the right panel of Fig.~\ref{EdGB_scalarCoupling}: by Eq.~(\ref{EdGBpoleValue}), $\varphi_A^{\rm pole}=\frac{1}{2}\ln\left(2\,x_{\rm pole}\,\mu_A^2/\alpha\right)$.

Qualitatively similar conclusions apply to EsGB theories with different coupling functions. Two interesting cases (quadratic and shift-symmetric theories) are discussed in Appendix~\ref{sec:AppQuadraticShiftSymmetric}.

\section{Conclusions}
\label{sec:conclusions}

The result we presented at the end of the previous section suggests that EdGB (and more generally, EsGB) theories must be treated with great care: when $\varphi_0$ is too large, the response of the BH to the scalar field diverges and the two-body problem is not even well-defined. Numerical work and/or higher-order expansions in the coupling seem necessary to verify this conclusion and to assess the convergence properties of the Pad\'e resummation.

However, the result seems compatible with hints from recent numerical work in various quadratic gravity theories. Simulations of stellar collapse and binary mergers have been successful in the decoupling limit~\cite{Benkel:2016rlz,Benkel:2016kcq,Okounkova:2017yby,Witek:2018dmd,Okounkova:2019dfo}, but the extension to the ``full'' theory presents notable conceptual and practical difficulties~\cite{Papallo:2017qvl,Cayuso:2017iqc,Allwright:2018rut,Ripley:2019hxt,Ripley:2019irj,Bernard:2019fjb}: for example, there are open sets of initial data for which the character of the system of equations changes from hyperbolic to elliptic in a compact region of the spacetime. Our work supports the expectation that quadratic theories should only be studied and trusted (in an effective field theory sense) in the weak-coupling regime.

An important limitation of our study is that the analytic expansion of our BH solutions (and hence their skeletonization) was performed around a Schwarzschild background. This rules out, by construction, the scalarized solutions discussed in the introduction. An extension of our work to scalarized solutions is necessary and important for gravitational-wave phenomenology.

At least in the small-$\alpha$ regime, the results of Ref.~\cite{Bernard:2018ivi}, supplemented by the Gauss-Bonnet contribution computed here [Eq.~(\ref{DeltaLAB})], yield the full EsGB Lagrangian at 3PN order. It will be interesting to extend the effective-one-body program to this more general class of theories: see~\cite{Julie:2017pkb,Julie:2017ucp} for similar work in ``ordinary'' scalar-tensor gravity, and~\cite{Khalil:2018aaj,Julie:2018lfp} for related work in EMD theory.

Our work should find application in analytical studies of dynamical scalarization (see~\cite{Khalil:2019wyy}) and in future studies of binary dynamics, using either the effective-one-body formalism or numerical relativity.

We wish to conclude by highlighting two technical results that we consider conceptually important:

\noindent
(1) At least during the inspiral, the mass function of skeletonized BHs [Eq.~(\ref{sensitivity})] is uniquely characterized by their Wald entropy. We conjecture that this might be true in all theories where the gravity sector differs from the Einstein-Hilbert Lagrangian. It will be interesting to test the validity of this conjecture and formally prove it.

\noindent
(2) The Gauss-Bonnet contributions to the fields are {\em finite} [Eq.~(\ref{h11})] and no regularization procedure is necessary, at least at 1PN order. While further work is necessary to determine whether this conclusion extends to higher PN orders, this intriguing result is yet another hint of the very special nature of EsGB gravity.\\

\noindent{\bf{\em Acknowledgments.}}
We thank Luc Blanchet, Thibault Damour, Nathalie Deruelle, Gilles Esposito-Far\`ese, Leonardo Gualtieri, Andrea Maselli, Nelson Merino, Hector O.~Silva, Leo Stein, Kent Yagi and Nico Yunes for discussions and suggestions, and the physics department at the University of Rome ``Sapienza'' for hospitality while this work was being completed.
We are particularly grateful to Luc Blanchet for sharing his expertise on the Fock (nonrelativistic) function (\ref{FockFunction}).
E.B. and F.L.J. are supported by NSF Grant No. PHY-1841464, NSF Grant No. AST-1841358, NSF-XSEDE Grant No. PHY-090003, and NASA ATP Grant No. 17-ATP17-0225.
The authors would like to acknowledge networking support by the GWverse COST Action 
CA16104, ``Black holes, gravitational waves and fundamental physics.''
We acknowledge support from the Amaldi Research Center funded by the MIUR program ``Dipartimento di Eccellenza''~(CUP: B81I18001170001).

\appendix

\section{The Einstein-scalar-Gauss-Bonnet field equations in arbitrary dimension\label{appFieldEqn}}

Let us generalize our vacuum action (\ref{vacuumAction}) to arbitrary dimensions:
\begin{equation}
I^D\!=\!\int\!\frac{d^Dx\sqrt{-g}}{16\pi}\bigg(\!R-2g^{\mu\nu}\partial_\mu\varphi\partial_\nu\varphi+\alpha f(\varphi)\mathcal R^2_{\rm GB}\!\bigg)\ .\label{vacuumActionD}
\end{equation}
In order to derive the associated Einstein field equations, it is useful to rewrite the Gauss-Bonnet scalar as \cite{Myers:1987yn, Deruelle:2018vtt}
\begin{align}
  \mathcal R_{\rm GB}^2=R^{\mu\nu\rho\sigma}P_{\mu\nu\rho\sigma}
\end{align}
with
\begin{align}
  P^{\mu\nu}_{\ \ \,\,\rho\sigma}&=R^{\mu\nu}_{\ \ \,\rho\sigma}-2\delta^\mu_{[\rho}R_{\sigma]}^\nu+2\delta^\nu_{[\rho}R_{\sigma]}^\mu+\delta^\mu_{[\rho}\delta_{\sigma]}^\nu R\\
&=\frac{1}{4}\delta^{\mu\nu\alpha_1\alpha_2}_{\rho\sigma\beta_1\beta_2}R_{\quad\ \,\alpha_1\alpha_2}^{\beta_1\beta_2}\ ,\nonumber
\end{align}
where $\delta^{\alpha_1\cdots\alpha_N}_{\beta_1\cdots\beta_N}$ denotes the generalized Kronecker symbol, which is the determinant of the $N\times N$ matrix $M$ built from ordinary Kronecker symbols as $M^i_j=\delta^{\alpha_i}_{\beta_j}$. The quantity $P_{\mu\nu\rho\sigma}$ has the symmetries of the Riemann tensor and is divergenceless: it can be easily shown using the Bianchi identities that $\nabla_\mu P^\mu_{\ \, \nu\rho\sigma}=0$.

The variation of the last term of (\ref{vacuumActionD}) with respect to $g^{\mu\nu}$ can therefore be written as:
\begin{align}
  \delta_{(g)}\!&\int \! d^Dx\sqrt{-g}f(\varphi)\mathcal R_{\rm GB}^2\label{variationGBarbitraryD}\\
  &=\int\!  d^Dx\sqrt{-g}f(\varphi)\left(H_{\mu\nu}\,\delta g^{\mu\nu}+2P_\mu^{\ \nu\rho\sigma}\delta R^\mu_{\ \nu\rho\sigma}\right)\ ,\nonumber
\end{align}
 where                                                                       
\begin{align}
H^\mu_\nu&= 2R^\mu_{\ \alpha\beta\gamma}P_{\nu}^{\ \alpha\beta\gamma}-\frac{1}{2}\delta^\mu_\nu\mathcal R_{\rm GB}^2\nonumber\\
&=-\frac{1}{8}\delta^{\mu\,\alpha_1\alpha_2\alpha_3\alpha_4}_{\,\nu\,\beta_1\beta_2\beta_3\beta_4}R_{\quad\ \,\alpha_1\alpha_2}^{\beta_1\beta_2}R_{\quad\ \,\alpha_3\alpha_4}^{\beta_3\beta_4}\nonumber
\end{align}
is the Gauss-Bonnet tensor. Now, using successively $\delta R^\mu_{\ \nu\rho\sigma}=2\nabla_{[\rho}\delta\Gamma^\mu_{\sigma]\nu}$ with $\delta \Gamma^\mu_{\nu\rho}=\frac{1}{2}g^{\mu\lambda}(\nabla_\nu\delta g_{\lambda\rho}+\nabla_\rho\delta g_{\lambda\nu}-\nabla_\lambda\delta g_{\nu\rho})$, integration by parts and the properties of $P_{\mu\nu\rho\sigma}$, one finds
\begin{align}
\delta_{(g)}\!&\int \! d^Dx\sqrt{-g}f(\varphi)\mathcal R_{\rm GB}^2\\
&=\int\! d^Dx\sqrt{-g}\left(f(\varphi) H_{\mu\nu}+4P_{\mu\alpha\nu\beta}\nabla^\alpha\nabla^\beta f(\varphi)\right)\delta g^{\mu\nu}\ ,\nonumber
\end{align}
modulo boundary terms ignored here.

The variation of the first two terms in (\ref{vacuumActionD}) is elementary, and the full Einstein field equations are thus, in any dimension $D$:
\begin{align}
R_{\mu\nu}-\frac{1}{2}g_{\mu\nu}R&=2\partial_\mu\varphi\partial_\nu\varphi-g_{\mu\nu}(\partial\varphi)^2\label{fieldEqnDimD}\\
&-\alpha \bigg(f(\varphi) H_{\mu\nu}+4 P_{\mu\alpha\nu\beta}\nabla^\alpha\nabla^\beta f(\varphi)\bigg)\ .\nonumber
\end{align}

When $D\leqslant 4$, the Gauss-Bonnet tensor $H_{\mu\nu}$ vanishes identically, as obvious from its expression above in terms of the rank-five generalized Kronecker symbol. Taking the trace of (\ref{fieldEqnDimD}) finally yields Eq.~(\ref{einsteinVacFieldEqn}).

\section{Einstein-scalar-Gauss-Bonnet black holes at order $\mathcal O(\epsilon^4)$\label{appendixFullBHsolution}}
Using the notation $f^{(n)}_\infty\equiv (d^n\!f/d\varphi^n)(\varphi_\infty)$ and recalling that $u=2m/r$, the remaining contributions to the static, spherically symmetric BH solutions (\ref{SSSansatz})--(\ref{BHsolution}) to the vacuum field equations (\ref{vacFieldEqn}) are:

\begin{widetext}
\begingroup\small
\Wider[0em]{
\begin{subequations}
\begin{align}
A_3=\frac{f^{\text{(2)}}_{\infty }}{f^{\text{(1)}}_{\infty
   }} \left(-\frac{73 u^3}{90} +\frac{73
   u^4}{36}+\frac{647 u^5}{450}+\frac{557 u^6}{900}+\frac{1189 u^7}{3150}-\frac{243
   u^8}{140}-\frac{667 u^9}{945}-\frac{43 u^{10}}{108}\right)\ ,
 \end{align}
 \begin{align}
 &A_4=\left(-\frac{73}{45}-\frac{362129 {f^{\text{(2)}}_{\infty }}^2}{226800
   {f^{\text{(1)}}_{\infty }}^2}-\frac{12511 f^{\text{(3)}}_{\infty }}{22680
   f^{\text{(1)}}_{\infty }}\right) u^3
   +\left(\frac{73}{18}+\frac{1139191
   {f^{\text{(2)}}_{\infty }}^2}{453600 {f^{\text{(1)}}_{\infty
   }}^2}+\frac{12511 f^{\text{(3)}}_{\infty }}{9072 f^{\text{(1)}}_{\infty }}\right)
   u^4
   +\left(-\frac{298}{225}+\frac{7993913 {f^{\text{(2)}}_{\infty
   }}^2}{2268000 {f^{\text{(1)}}_{\infty }}^2}-\frac{12511
   f^{\text{(3)}}_{\infty }}{113400 f^{\text{(1)}}_{\infty }}\right)
   u^5\nonumber\\
&+\left(\frac{439}{150}+\frac{1694561 {f^{\text{(2)}}_{\infty
   }}^2}{1134000 {f^{\text{(1)}}_{\infty }}^2}+\frac{138689
   f^{\text{(3)}}_{\infty }}{226800 f^{\text{(1)}}_{\infty }}\right)
   u^6
   +\left(-\frac{2231}{450}+\frac{1425247 {f^{\text{(2)}}_{\infty
   }}^2}{1587600 {f^{\text{(1)}}_{\infty }}^2}+\frac{218069
   f^{\text{(3)}}_{\infty }}{396900 f^{\text{(1)}}_{\infty }}\right)
   u^7
   +\left(\frac{9979}{2520}-\frac{11507039 {f^{\text{(2)}}_{\infty
   }}^2}{6350400 {f^{\text{(1)}}_{\infty }}^2}+\frac{288377
   f^{\text{(3)}}_{\infty }}{635040 f^{\text{(1)}}_{\infty }}\right)
   u^8\nonumber\\
&+\left(-\frac{443}{280}-\frac{27378403 {f^{\text{(2)}}_{\infty
   }}^2}{19051200 {f^{\text{(1)}}_{\infty }}^2}-\frac{132829
   f^{\text{(3)}}_{\infty }}{238140 f^{\text{(1)}}_{\infty }}\right)
   u^9
   +\left(\frac{8203}{450}-\frac{169633 {f^{\text{(2)}}_{\infty
   }}^2}{170100 {f^{\text{(1)}}_{\infty }}^2}-\frac{150041
   f^{\text{(3)}}_{\infty }}{272160 f^{\text{(1)}}_{\infty }}\right)
   u^{10}
   +\left(-\frac{779}{330}-\frac{13558757 {f^{\text{(2)}}_{\infty
   }}^2}{18711000 {f^{\text{(1)}}_{\infty }}^2}-\frac{354643
   f^{\text{(3)}}_{\infty }}{748440 f^{\text{(1)}}_{\infty }}\right)
   u^{11}\ \ \nonumber\\
&+\left(-\frac{7}{5}-\frac{16763 {f^{\text{(2)}}_{\infty }}^2}{81000
   {f^{\text{(1)}}_{\infty }}^2}-\frac{493 f^{\text{(3)}}_{\infty }}{3240
   f^{\text{(1)}}_{\infty }}\right) u^{12}
   +\left(-\frac{9908}{825}-\frac{5
   {f^{\text{(2)}}_{\infty }}^2}{88 {f^{\text{(1)}}_{\infty
   }}^2}-\frac{f^{\text{(3)}}_{\infty }}{22 f^{\text{(1)}}_{\infty }}\right) u^{13}\ ,\hspace*{3cm}\\
   \nonumber
\end{align}\label{BHsolutionA}%
\end{subequations}%
}
\endgroup
\end{widetext}

\begin{widetext}
\begingroup\small
\Wider[0em]{
\begin{subequations}
\begin{align}
B_3=-\frac{f^{\text{(2)}}_{\infty }}{f^{\text{(1)}}_{\infty }} \left(\frac{73 u^2}{30}+\frac{73
   u^3}{45}+\frac{73 u^4}{36}+\frac{103 u^5}{50}+\frac{413 u^6}{225}+\frac{57
   u^7}{35}+\frac{253 u^8}{420}+\frac{11 u^9}{54}\right)\ ,
 \end{align}
 \begin{align}
&B_4=-\left(\frac{73}{15}+\frac{362129 {f^{\text{(2)}}_{\infty }}^2}{75600
   {f^{\text{(1)}}_{\infty }}^2}+\frac{12511 f^{\text{(3)}}_{\infty }}{7560
   f^{\text{(1)}}_{\infty }}\right) u^2
   -\left(\frac{146}{45}+\frac{362129
   {f^{\text{(2)}}_{\infty }}^2}{113400 {f^{\text{(1)}}_{\infty
   }}^2}+\frac{12511 f^{\text{(3)}}_{\infty }}{11340 f^{\text{(1)}}_{\infty }}\right)
   u^3
   -\left(\frac{73}{18}+\frac{1586827 {f^{\text{(2)}}_{\infty }}^2}{453600
   {f^{\text{(1)}}_{\infty }}^2}+\frac{12511 f^{\text{(3)}}_{\infty }}{9072
   f^{\text{(1)}}_{\infty }}\right) u^4\nonumber\\
&-\left(\frac{169}{75}+\frac{254393
   {f^{\text{(2)}}_{\infty }}^2}{63000 {f^{\text{(1)}}_{\infty
   }}^2}+\frac{12511 f^{\text{(3)}}_{\infty }}{12600 f^{\text{(1)}}_{\infty }}\right)
   u^5
   -\left(\frac{847}{450}+\frac{121219 {f^{\text{(2)}}_{\infty }}^2}{32400
   {f^{\text{(1)}}_{\infty }}^2}+\frac{16831 f^{\text{(3)}}_{\infty }}{16200
   f^{\text{(1)}}_{\infty }}\right) u^6
   -\left(\frac{1}{45}+\frac{2691779
   {f^{\text{(2)}}_{\infty }}^2}{793800 {f^{\text{(1)}}_{\infty
   }}^2}+\frac{21394 f^{\text{(3)}}_{\infty }}{19845 f^{\text{(1)}}_{\infty }}\right)
   u^7\nonumber\\
&-\left(\frac{2549}{2520}+\frac{479659 {f^{\text{(2)}}_{\infty
   }}^2}{235200 {f^{\text{(1)}}_{\infty }}^2}+\frac{25783
   f^{\text{(3)}}_{\infty }}{23520 f^{\text{(1)}}_{\infty }}\right)
   u^8
   -\left(\frac{11}{45}+\frac{94471 {f^{\text{(2)}}_{\infty }}^2}{85050
   {f^{\text{(1)}}_{\infty }}^2}+\frac{18829 f^{\text{(3)}}_{\infty }}{27216
   f^{\text{(1)}}_{\infty }}\right) u^9
   -\left(\frac{583}{90}+\frac{35633
   {f^{\text{(2)}}_{\infty }}^2}{68040 {f^{\text{(1)}}_{\infty
   }}^2}+\frac{6079 f^{\text{(3)}}_{\infty }}{17010 f^{\text{(1)}}_{\infty }}\right)
   u^{10}\nonumber\\
&-\left(\frac{4504}{825}+\frac{5089 {f^{\text{(2)}}_{\infty
   }}^2}{37125 {f^{\text{(1)}}_{\infty }}^2}+\frac{611 f^{\text{(3)}}_{\infty
   }}{5940 f^{\text{(1)}}_{\infty }}\right) u^{11}
   -\left(\frac{1329}{275}+\frac{205
   {f^{\text{(2)}}_{\infty }}^2}{7128 {^{\text{(1)}}_{\infty
   }}^2}+\frac{41 f^{\text{(3)}}_{\infty }}{1782 f^{\text{(1)}}_{\infty }}\right)
   u^{12}\ ,\\
   \nonumber
\end{align}\label{BHsolutionB}%
\end{subequations}
}
\endgroup
\end{widetext}
and finally
\begin{widetext}
\begingroup\small
\begin{subequations}
\Wider[0em]{
\begin{align}
  \varphi_2=\frac{f^{(2)}_\infty}{f^{(1)}_\infty}\left[\frac{73}{60}\left(u+\frac{u^2}{2}+\frac{u^3}{3}+\frac{u^4}{4}\right)+\frac{7u^5}{75}+\frac{u^6}{36}\right]\ ,
\end{align}
\begin{align}
&\varphi_3=\left(\frac{73}{30}+\frac{12511 {f^{\text{(2)}}_{\infty }}^2}{7560
   {f^{\text{(1)}}_{\infty }}^2}+\frac{12511 f^{\text{(3)}}_{\infty }}{15120
   f^{\text{(1)}}_{\infty }}\right) u
 +\left(\frac{73}{60}+\frac{12511
   {f^{\text{(2)}}_{\infty }}^2}{15120 {f^{\text{(1)}}_{\infty
   }}^2}+\frac{12511 f^{\text{(3)}}_{\infty }}{30240 f^{\text{(1)}}_{\infty }}\right)
   u^2
   +\left(\frac{103}{90}+\frac{12511 {f^{\text{(2)}}_{\infty }}^2}{22680
   {f^{\text{(1)}}_{\infty }}^2}+\frac{12511 f^{\text{(3)}}_{\infty }}{45360
   f^{\text{(1)}}_{\infty }}\right) u^3\nonumber\\   
   &+\left(\frac{133}{120}+\frac{12511
   {f^{\text{(2)}}_{\infty }}^2}{30240 {f^{\text{(1)}}_{\infty
   }}^2}+\frac{12511 f^{\text{(3)}}_{\infty }}{60480 f^{\text{(1)}}_{\infty }}\right)
   u^4
   +\left(\frac{51}{50}+\frac{449 {f^{\text{(2)}}_{\infty }}^2}{3024
   {f^{\text{(1)}}_{\infty }}^2}+\frac{12511 f^{\text{(3)}}_{\infty }}{75600
   f^{\text{(1)}}_{\infty }}\right) u^5
   +\left(\frac{73}{180}+\frac{28531
   {f^{\text{(2)}}_{\infty }}^2}{453600 {f^{\text{(1)}}_{\infty
   }}^2}+\frac{1595 f^{\text{(3)}}_{\infty }}{18144 f^{\text{(1)}}_{\infty }}\right)
   u^6\nonumber\\
   &+\left(\frac{17}{10}+\frac{13201 {f^{\text{(2)}}_{\infty }}^2}{529200
   {f^{\text{(1)}}_{\infty }}^2}+\frac{839 f^{\text{(3)}}_{\infty }}{21168
   f^{\text{(1)}}_{\infty }}\right) u^7
   +\left(\frac{57}{40}+\frac{239
   {f^{\text{(2)}}_{\infty }}^2}{43200 {f^{\text{(1)}}_{\infty
   }}^2}+\frac{35 f^{\text{(3)}}_{\infty }}{3456 f^{\text{(1)}}_{\infty }}\right)
   u^8
   +\left(\frac{173}{135}+\frac{{f^{\text{(2)}}_{\infty }}^2}{972
   {f^{\text{(1)}}_{\infty }}^2}+\frac{f^{\text{(3)}}_{\infty }}{486
   f^{\text{(1)}}_{\infty }}\right) u^9\ ,
\end{align}
}
\Wider[2.5em]{
\begin{align}
&\varphi_4=\frac{f^{\text{(2)}}_{\infty }}{f^{\text{(1)}}_{\infty }} \Bigg[\left(\frac{143467}{8316}+\frac{227192473
   {f^{\text{(2)}}_{\infty }}^2}{99792000 {f^{\text{(1)}}_{\infty
   }}^2}+\frac{31557593 f^{\text{(3)}}_{\infty }}{9072000 f^{\text{(1)}}_{\infty
   }}+\frac{799607 f^{\text{(4)}}_{\infty }}{1995840 f^{\text{(2)}}_{\infty }}\right)u
   +\left(\frac{143467}{16632}+\frac{227192473 {f^{\text{(2)}}_{\infty
   }}^2}{199584000 {f^{\text{(1)}}_{\infty }}^2}+\frac{31557593
   f^{\text{(3)}}_{\infty }}{18144000 f^{\text{(1)}}_{\infty }}+\frac{799607
   f^{\text{(4)}}_{\infty }}{3991680 f^{\text{(2)}}_{\infty }}\right)
   u^2\nonumber\\
&+\left(\frac{434551}{62370}+\frac{227192473 {f^{\text{(2)}}_{\infty
   }}^2}{299376000 {f^{\text{(1)}}_{\infty }}^2}+\frac{31557593
   f^{\text{(3)}}_{\infty }}{27216000 f^{\text{(1)}}_{\infty }}+\frac{799607
   f^{\text{(4)}}_{\infty }}{5987520 f^{\text{(2)}}_{\infty }}\right)
   u^3
   +\left(\frac{1020869}{166320}+\frac{227192473 {f^{\text{(2)}}_{\infty
   }}^2}{399168000 {f^{\text{(1)}}_{\infty }}^2}+\frac{31557593
   f^{\text{(3)}}_{\infty }}{36288000 f^{\text{(1)}}_{\infty }}+\frac{799607
   f^{\text{(4)}}_{\infty }}{7983360 f^{\text{(2)}}_{\infty }}\right)
   u^4\nonumber\\
&+\left(\frac{2126053}{415800}+\frac{14761939 {f^{\text{(2)}}_{\infty}}^2}{71280000 {f^{\text{(1)}}_{\infty }}^2}+\frac{3703949
   f^{\text{(3)}}_{\infty }}{6480000 f^{\text{(1)}}_{\infty }}+\frac{799607
   f^{\text{(4)}}_{\infty }}{9979200 f^{\text{(2)}}_{\infty }}\right)
   u^5
+\left(\frac{8051381}{2494800}+\frac{53790013 {f^{\text{(2)}}_{\infty
   }}^2}{598752000 {f^{\text{(1)}}_{\infty }}^2}+\frac{17053103
   f^{\text{(3)}}_{\infty }}{54432000 f^{\text{(1)}}_{\infty }}+\frac{799607
   f^{\text{(4)}}_{\infty }}{11975040 f^{\text{(2)}}_{\infty }}\right)
   u^6\nonumber\\
&+\left(\frac{2128363}{582120}+\frac{178679 {f^{\text{(2)}}_{\infty
   }}^2}{4752000 {f^{\text{(1)}}_{\infty }}^2}+\frac{1469029
   f^{\text{(3)}}_{\infty }}{9072000 f^{\text{(1)}}_{\infty }}+\frac{633287
   f^{\text{(4)}}_{\infty }}{13970880 f^{\text{(2)}}_{\infty }}\right)
   u^7
   +\left(\frac{85573}{16632}+\frac{20000597 {f^{\text{(2)}}_{\infty
   }}^2}{1862784000 {f^{\text{(1)}}_{\infty }}^2}+\frac{35999071
   f^{\text{(3)}}_{\infty }}{508032000 f^{\text{(1)}}_{\infty }}+\frac{59921
   f^{\text{(4)}}_{\infty }}{2280960 f^{\text{(2)}}_{\infty }}\right)
   u^8\nonumber\\
&+\left(\frac{4017613}{748440}+\frac{3517861 {f^{\text{(2)}}_{\infty
   }}^2}{1047816000 {f^{\text{(1)}}_{\infty }}^2}+\frac{15156781
   f^{\text{(3)}}_{\infty }}{571536000 f^{\text{(1)}}_{\infty }}+\frac{449
   f^{\text{(4)}}_{\infty }}{40095 f^{\text{(2)}}_{\infty }}\right)
   u^9
   +\left(\frac{22617977}{4158000}+\frac{9691879 {f^{\text{(2)}}_{\infty
   }}^2}{10478160000 {f^{\text{(1)}}_{\infty }}^2}+\frac{15718103
   f^{\text{(3)}}_{\infty }}{1905120000 f^{\text{(1)}}_{\infty }}+\frac{2729
   f^{\text{(4)}}_{\infty }}{712800 f^{\text{(2)}}_{\infty }}\right)
   u^{10}\qquad\nonumber\\
&+\left(\frac{9714977}{4573800}+\frac{7553 {f^{\text{(2)}}_{\infty
   }}^2}{47044800 {f^{\text{(1)}}_{\infty }}^2}+\frac{13891
   f^{\text{(3)}}_{\infty }}{8553600 f^{\text{(1)}}_{\infty }}+\frac{65
   f^{\text{(4)}}_{\infty }}{78408 f^{\text{(2)}}_{\infty }}\right)
   u^{11}
   +\left(\frac{2447}{3240}+\frac{{f^{\text{(2)}}_{\infty }}^2}{46656
   {f^{\text{(1)}}_{\infty }}^2}+\frac{11 f^{\text{(3)}}_{\infty }}{46656
   f^{\text{(1)}}_{\infty }}+\frac{f^{\text{(4)}}_{\infty }}{7776
   f^{\text{(2)}}_{\infty }}\right) u^{12}\Bigg]\ . \\ \nonumber
\end{align}
}\label{BHsolutionPhiApp}
\end{subequations}
\endgroup
\end{widetext}

It is then simple to compute the Kretschmann scalar of the spacetime, with the result (recall that $\epsilon=\alpha f^{(1)}_\infty/4m^2$):
\begin{widetext}
\begingroup\small
\Wider[2.5em]{
\begin{align}
  &\hspace*{4cm} R^{\mu\nu\rho\sigma}R_{\mu\nu\rho\sigma}=\frac{1}{m^4}\Bigg[\frac{3 u^6}{4}+\epsilon ^2 \left(-u^7+2 u^8-\frac{33 u^9}{2}+\frac{7
    u^{10}}{4}+u^{11}+\frac{138 u^{12}}{5}\right)
  \label{eq:Kretschmann}\\
&+\epsilon ^3 \left(-\frac{73
   f^{\text{(2)}}_{\infty } }{30 f^{\text{(1)}}_{\infty }}u^7+\frac{73
   f^{\text{(2)}}_{\infty }}{15 f^{\text{(1)}}_{\infty }} u^8-\frac{73
   f^{\text{(2)}}_{\infty }}{4 f^{\text{(1)}}_{\infty }} u^9-\frac{347
   f^{\text{(2)}}_{\infty }}{20 f^{\text{(1)}}_{\infty
   }} u^{10}-\frac{1799 f^{\text{(2)}}_{\infty }}{200
   f^{\text{(1)}}_{\infty }} u^{11}-\frac{1013 f^{\text{(2)}}_{\infty }
   }{150 f^{\text{(1)}}_{\infty }}u^{12}+\frac{5687 f^{\text{(2)}}_{\infty
   }}{105 f^{\text{(1)}}_{\infty }} u^{13}+\frac{1133
   f^{\text{(2)}}_{\infty }}{42 f^{\text{(1)}}_{\infty
   }} u^{14}+\frac{1309 f^{\text{(2)}}_{\infty } }{72
   f^{\text{(1)}}_{\infty }}u^{15}\right)\quad\quad\quad\nonumber\\
&+\epsilon ^4 \Bigg(\left(-\frac{73}{15}-\frac{362129
   {f^{\text{(2)}}_{\infty }}^2}{75600 {f^{\text{(1)}}_{\infty
   }}^2}-\frac{12511 f^{\text{(3)}}_{\infty }}{7560 f^{\text{(1)}}_{\infty }}\right)
   u^7
   +\left(\frac{629}{60}+\frac{362129 {f^{\text{(2)}}_{\infty
   }}^2}{37800 {f^{\text{(1)}}_{\infty }}^2}+\frac{12511
   f^{\text{(3)}}_{\infty }}{3780 f^{\text{(1)}}_{\infty }}\right)
   u^8
   +\left(-39-\frac{1139191 {f^{\text{(2)}}_{\infty }}^2}{50400
   {f^{\text{(1)}}_{\infty }}^2}-\frac{12511 f^{\text{(3)}}_{\infty }}{1008
   f^{\text{(1)}}_{\infty }}\right) u^9\nonumber\\
&+\left(\frac{2023}{60}-\frac{59519
   {f^{\text{(2)}}_{\infty }}^2}{1344 {f^{\text{(1)}}_{\infty
   }}^2}+\frac{12511 f^{\text{(3)}}_{\infty }}{5040 f^{\text{(1)}}_{\infty }}\right)
   u^{10}
   +\left(-\frac{23057}{300}-\frac{276017 {f^{\text{(2)}}_{\infty
   }}^2}{12000 {f^{\text{(1)}}_{\infty }}^2}-\frac{73889
   f^{\text{(3)}}_{\infty }}{7200 f^{\text{(1)}}_{\infty }}\right)
   u^{11}
   +\left(\frac{5769}{25}-\frac{1288367 {f^{\text{(2)}}_{\infty
   }}^2}{75600 {f^{\text{(1)}}_{\infty }}^2}-\frac{56587
   f^{\text{(3)}}_{\infty }}{4725 f^{\text{(1)}}_{\infty }}\right)
   u^{12}\nonumber\\
&+\left(-\frac{16417}{105}+\frac{6878401 {f^{\text{(2)}}_{\infty
   }}^2}{117600 {f^{\text{(1)}}_{\infty }}^2}-\frac{146081
   f^{\text{(3)}}_{\infty }}{11760 f^{\text{(1)}}_{\infty }}\right)
   u^{13}
   +\left(\frac{77503}{840}+\frac{78844487 {f^{\text{(2)}}_{\infty
   }}^2}{1411200 {f^{\text{(1)}}_{\infty }}^2}+\frac{77767
   f^{\text{(3)}}_{\infty }}{3528 f^{\text{(1)}}_{\infty }}\right)
   u^{14}
   +\left(-\frac{14417}{12}+\frac{655748 {f^{\text{(2)}}_{\infty
   }}^2}{14175 {f^{\text{(1)}}_{\infty }}^2}+\frac{931387
   f^{\text{(3)}}_{\infty }}{36288 f^{\text{(1)}}_{\infty }}\right)
   u^{15}\qquad\quad \nonumber\\
&+\left(\frac{9733}{60}+\frac{7468747 {f^{\text{(2)}}_{\infty
   }}^2}{189000 {f^{\text{(1)}}_{\infty }}^2}+\frac{24436
   f^{\text{(3)}}_{\infty }}{945 f^{\text{(1)}}_{\infty }}\right)
   u^{16}
   +\left(\frac{1051}{10}+\frac{356657 {f^{\text{(2)}}_{\infty
   }}^2}{27000 {f^{\text{(1)}}_{\infty }}^2}+\frac{2623
   f^{\text{(3)}}_{\infty }}{270 f^{\text{(1)}}_{\infty }}\right)
   u^{17}
   +\left(\frac{4028179}{3300}+\frac{19915 {f^{\text{(2)}}_{\infty
   }}^2}{4752 {f^{\text{(1)}}_{\infty }}^2}+\frac{3983 f^{\text{(3)}}_{\infty
   }}{1188 f^{\text{(1)}}_{\infty }}\right) u^{18}\Bigg)\Bigg]\,.\nonumber\\
   \nonumber
\end{align}
}
\endgroup
\end{widetext}
This expression diverges only at the origin $r=0$, showing that our BH solution is regular everywhere outside the horizon, since $r_{\rm H}>0$ when $\epsilon\ll 1$: cf. Eq.~(\ref{horizonLocation}).

\section{The thermodynamical variables of Einstein-scalar-Gauss-Bonnet black holes\label{appThermoParameters}}

The solution presented in Sec.~\ref{sectionTNsolution} and Appendix \ref{appendixFullBHsolution} can be characterized by the thermodynamic quantities defined in Sec.~\ref{sectionThermo}. Here, we give their expressions in terms of the integration constants $m$ and $\varphi_\infty$, denoting $\epsilon\equiv \alpha f^{(1)}_\infty/4m^2$.

The location of the horizon $u_{\rm H}=2m/r_{\rm H}$, the temperature $T$ (\ref{temperature}), the Wald entropy $S_{\rm w}$ -- as defined in Eqs.~(\ref{waldEntropy1}) and (\ref{waldEntropy2}) -- and the scalar ``charge'' $D$ defined below (\ref{katzMass}) are given by

\begin{widetext}
\begin{align}
u_{\rm H}&=1+\frac{\epsilon^2}{3}+\epsilon^3\frac{73 f^{(2)}_\infty}{90 f^{(1)}_\infty}+\epsilon^4\left(\frac{1646}{495}+\frac{362129 {f^{(2)}_\infty}^2}{226800 {f^{(1)}_\infty}^2}+\frac{12511 f^{(3)}_\infty}{22680 f^{(1)}_\infty}\right)+\mathcal O(\epsilon^5)\ .\label{horizonLocation}
\\
  T&=8\pi m \left[1+\epsilon ^2\frac{73 }{30}+\epsilon ^3\frac{12511 f^{(2)}_\infty }{1890 f^{(1)}_\infty}+\epsilon ^4 \left(\frac{4010597}{138600}+\frac{227192473 {f^{(2)}_\infty}^2}{16632000 {f^{(1)}_\infty}^2}+\frac{799607 f^{(3)}_\infty}{166320 f^{(1)}_\infty}\right)+\mathcal O(\epsilon^5)\right]\ .\label{temperatureMPHI}
\\
  S_{\rm w}&=4\pi m^2\left[1 + \epsilon\frac{4f_\infty}{f^{(1)}_\infty}+\epsilon^2\frac{73}{30} +\epsilon^3\frac{12511 f^{(2)}_\infty}{3780 f^{(1)}_\infty}\right.  \left.+\epsilon^4\left(\frac{3189931}{415800}+\frac{227192473 {f^{(2)}_\infty}^2}{49896000 {f^{(1)}_\infty}^2} + \frac{
      799607 f^{(3)}_\infty}{498960 f^{(1)}_\infty}\right)+\mathcal O(\epsilon^5)\right]\ . \label{entropyMPHI} 
\\
  D&= 2m  \left[\epsilon+ \epsilon^2 \frac{73 f^{(2)}_\infty}{60 f^{(1)}_\infty}+\epsilon ^3 \left(\frac{73}{30}+\frac{12511 {f^{(2)}_\infty}^2}{7560 {f^{(1)}_\infty}^2}+\frac{12511 f^{(3)}_\infty}{15120 f^{(1)}_\infty}\right)\right. \nonumber \\
&\left.+\epsilon ^4 \left(\frac{143467 }{8316}+\frac{227192473 {f^{(2)}_\infty}^2}{99792000 {f^{(1)}_\infty}^2}+\frac{31557593  f^{(3)}_\infty}{9072000 f^{(1)}_\infty}+\frac{799607 f^{(4)}_\infty}{1995840 f^{(2)}_\infty}\right)\frac{f^{(2)}_\infty}{f^{(1)}_\infty}+\mathcal O(\epsilon^5)\right]\ . \label{scalarChargeMPHI}
\end{align}
\end{widetext}

\section{The two-body Lagrangian at first post-Newtonian order\label{AppendixPNLagrangian}}
In this appendix we derive the PN two-body Lagrangian of EsGB theories, Eq.~(\ref{1PNLagrangian}). For a bound binary system, we compute the relativistic  corrections in the weak field, slow velocity approximation at order $\mathcal O(m/r)\sim\mathcal O(v^2)$, where $r$ is the distance separating the bodies and $v$ is their relative velocity.

Our first goal is to solve the EsGB field equations (\ref{fieldEqnSkel2}) sourced by two point particles:
\begin{widetext}
\begin{subequations}
\begin{align}
R_{\mu\nu}&=2\partial_\mu\varphi\partial_\nu\varphi-4\alpha\left(P_{\mu\alpha\nu\beta}-\frac{1}{2}g_{\mu\nu}P_{\alpha\beta}\right)\nabla^\alpha\nabla^\beta f(\varphi)+8\pi\sum_A\left(T^A_{\mu\nu}-\frac{1}{2}g_{\mu\nu}T^A\right)\ ,\label{einsteinFieldEqnAppendix}\\
\Box\varphi&=-\frac{1}{4}\alpha f'(\varphi)\mathcal R_{\rm GB}^2+4\pi\sum_A\frac{ds_A}{dt}\frac{dm_A}{d\varphi}\frac{\delta^{(3)}(\mathbf{x}-\mathbf x_A(t))}{\sqrt{-g}}\ ,\label{KleinGordonFieldEqnAppendix}
\end{align}
\label{fieldEqnAppendix}%
\end{subequations}
\end{widetext}
 where we recall that $T_A^{\mu\nu}$ is the distributional stress-energy tensor of the skeletonized body $A$ located at $x_A^\mu=(t,\mathbf x_A)$:
\begin{equation}
T_A^{\mu\nu}=m_A(\varphi)\frac{\delta^{(3)}(\mathbf{x}-\mathbf x_A(t))}{\sqrt{gg_{\alpha\beta}\frac{dx_A^\alpha}{dt}\frac{dx_A^\beta}{dt}}}\frac{dx_A^\mu}{dt}\frac{dx_A^\nu}{dt}\ .
\end{equation}

At 1PN order and in Cartesian coordinates, it is convenient to expand the metric around Minkowski as~\cite{Damour:1990pi}:
\begin{subequations}
\begin{align}
g_{00}&=-e^{-2U}+\mathcal O(v^6)\ ,\\
g_{0i}&=-4g_i+\mathcal O(v^5)\ ,\\
g_{ij}&=\delta_{ij}e^{2U}+\mathcal O(v^4)\ ,
\end{align}
\end{subequations}
where, as we show below, $U=\mathcal O(v^2)$ and $g_i=\mathcal O(v^3)$. We can also expand the scalar field $\varphi$ as
\begin{equation}
\varphi=\varphi_0+\delta\varphi+\mathcal O(v^6)\ ,
\end{equation}
with $\delta\varphi=\mathcal O(v^2)$. The masses $m_A(\varphi)$ are expanded around the value $\varphi_0$ of the scalar field at infinity, using the quantities defined in Eqs.~(\ref{DefAlpha}) and (\ref{DefBeta}):
\begin{equation}
m_A(\varphi)=m_A^0\left[1+\alpha_A^0\delta\varphi+\frac{1}{2}({\alpha_A^0}^2+\beta_A^0)\delta\varphi^2+\mathcal O(v^6)\right]\,.
\end{equation}
Here a ``$0$'' subscript indicates that the quantity is evaluated at $\varphi=\varphi_0$.
In a harmonic coordinate system and at 1PN order we have $\partial_\mu(\sqrt{-g}g^{\mu\nu})=\partial_t U+\partial_i g_i=0$,  $R^{00}=-\Box_\eta U+\mathcal O(v^6)$, and $R^{0i}=-2\Delta g_i+\mathcal O(v^5)$. The Gauss-Bonnet term contributes to the field equations through $P_{0i0j}+\frac{1}{2}P_{ij}=-(\partial_{ij}U)+\delta_{ij}\Delta U+\mathcal O(v^4)$ and $\mathcal R_{\rm GB}^2=8\left[(\partial_{ij}U)(\partial_{ij}U)-\Delta U\Delta U\right]+\mathcal O(v^6)$, so the field equations read
\begin{widetext}
\begin{subequations}
\begin{align}
  \Box_\eta U&=-4\pi\sum_A m_A^0\left[1+\frac{3}{2}\mathbf v_A^2-U+\alpha_A^0\delta\varphi\right]\delta^{(3)}(\mathbf x-\mathbf x_A(t))
  +4\alpha f'(\varphi_0)\left[\Delta\varphi\Delta U-(\partial_{ij}\varphi)(\partial_{ij}U)\right]+\mathcal O(v^6)\ ,\label{fieldEqnUappendix}\\
\Delta g_i&=-4\pi\sum_A m_A^0 v_A^i\delta^{(3)}(\mathbf x-\mathbf x_A(t))+\mathcal O(v^5)\ ,\\
\Box_\eta\varphi&=4\pi\sum_A m_A^0\alpha_A^0\left[1-\frac{1}{2}\mathbf v_A^2-U+\left(\alpha_A^0+\frac{\beta_A^0}{\alpha_A^0}\right)\delta\varphi\right]\delta^{(3)}(\mathbf x-\mathbf x_A(t))
+2\alpha f'(\varphi_0)\left[(\Delta U)^2-(\partial_{ij}U)(\partial_{ij}U)\right]+\mathcal O(v^6)\ , \label{fieldEqnPhiappendix}
\end{align}\label{fieldEqn1PN}
\end{subequations}
\end{widetext}
 where $\Box_\eta=\eta^{\mu\nu}\partial_\mu\partial_\nu$ is the flat D'Alembertian and $\Delta=\delta^{ij}\partial_i\partial_j$ is the flat Laplacian.

When the Gauss-Bonnet coupling is switched off, i.e., $\alpha=0$, the system above reduces to the standard scalar-tensor field equations at 1PN. We can now solve these equations using standard methods (see, e.g. \cite{Damour:1992we,Damour:1995kt} or \cite{Julie:2017rpw}) through the relativistic Green's function
\begin{equation}
\Box_\eta G(x,x')\equiv-4\pi\delta^{(3)}(\mathbf{x}-\mathbf{x'})\delta(t-t')\,,
\end{equation}
which, as we focus here on the conservative sector, is half-retarded, half-advanced:
\begin{align}
G(x,x')&=\frac{1}{2}\left[\frac{\delta(t-t'-|\mathbf{ x}-\mathbf{ x'}|}{|\mathbf{x}-\mathbf{x'}|}+\frac{\delta(t-t'+|\mathbf{x}-\mathbf{ x'}|}{|\mathbf{ x}-\mathbf{x'}|}\right]\nonumber\\
       &=\frac{\delta(t-t')}{|\mathbf{ x}-\mathbf{ x'}|}+\frac{|\mathbf{x}-\mathbf{ x'}|}{2}\partial_t^2\delta(t-t')+\cdots\,.
\end{align}
All derivatives are understood in a distributional sense.
 
The new $\alpha$-driven sources of Eqs.~(\ref{fieldEqnUappendix}) and (\ref{fieldEqnPhiappendix})  enter (formally)
at 1PN level. To evaluate them we must replace $U$ and $\varphi$ by their leading (0PN) expressions, yielding equations of the form:
\begin{align}
\Delta h_{12}=\Delta\frac{1}{|\mathbf x-\mathbf y_1|}\Delta\frac{1}{|\mathbf{ x}-\mathbf y_2|}-\partial_{ij}\frac{1}{|\mathbf{ x}-\mathbf y_1|}\partial_{ij}\frac{1}{|\mathbf{ x}-\mathbf y_2|}\ .\label{defGreensGB}
\end{align}
The solution $h_{12}(\mathbf{ x})$ can be found as follows. When $\mathbf y_1\neq\mathbf y_2$, we can replace  the gradients $\partial_i$ by derivatives with respect to the source locations $\mathbf y_1$ and $\mathbf y_2$:
\begin{align}
  \Delta h_{12}&=\left(\frac{\partial^2}{\partial y_1^i\partial y_1^i}\frac{\partial^2}{\partial y_2^j\partial y_2^j}-\frac{\partial^2}{\partial y_1^i\partial y_2^i}\frac{\partial^2}{\partial y_1^j\partial y_2^j}\right)\nonumber\\
                 &\times \frac{1}{|\mathbf{ x}-\mathbf y_1||\mathbf{ x}-\mathbf y_2|}\ .\label{trickBlanchet}
\end{align}
Now note that $\Delta^{-1}$ commutes with the $y^i$-derivatives, and recall the well-known result first established by Fock (see, e.g. \cite{Blanchet:2003gy}),
\begin{equation}
\Delta^{-1}\left(\frac{1}{|\mathbf{ x}-\mathbf y_1||\mathbf{ x}-\mathbf y_2|}\right)=\ln(|\mathbf{ x}-\mathbf y_1|+|\mathbf{ x}-\mathbf y_2|+|\mathbf y_1-\mathbf y_2|)\,.\label{FockFunction}
\end{equation}
A rather lengthy but straightforward calculation then yields:
\begin{widetext}
\begin{align}
  h_{12}(\mathbf{ x})&=\frac{1}{4|\mathbf{ x}-\mathbf y_1|^3|\mathbf{ x}-\mathbf y_2|^3}
                \left(|\mathbf{ x}-\mathbf y_1|^2+|\mathbf{ x}-\mathbf y_2|^2-|\mathbf y_1-\mathbf y_2|^2+\frac{|\mathbf{ x}-\mathbf y_1|^3+|\mathbf{ x}-\mathbf y_2|^3}{|\mathbf y_1-\mathbf y_2|}\right. \nonumber\\
                &+\left.\frac{|\mathbf{ x}-\mathbf y_1|^3|\mathbf{ x}-\mathbf y_2|^2+|\mathbf{ x}-\mathbf y_1|^2|\mathbf{ x}-\mathbf y_2|^3-|\mathbf{ x}-\mathbf y_1|^5-|\mathbf{ x}-\mathbf y_1|^5}{|\mathbf y_1-\mathbf y_2|^3}\right)\ . \label{h12}
\end{align}
\end{widetext}
It can be checked that the contribution from the first set of derivatives in Eq.~(\ref{trickBlanchet}) vanishes identically: the first, ``Dirac squared'' term in Eq.~(\ref{defGreensGB}) can be ignored.

The case $\mathbf y_1=\mathbf y_2$ can be inferred from Eq.~(\ref{h12}). Denoting
 $\mathbf n_1\equiv\frac{\mathbf{ x}-\mathbf y_1}{|\mathbf{ x}-\mathbf y_1|}$ and $\mathbf n_{12}\equiv\frac{\mathbf y_2-\mathbf y_1}{|\mathbf y_2-\mathbf y_1|}$, and taking the limit $\epsilon\equiv|\mathbf y_1-\mathbf y_2|\to 0$, we find:
\begin{align}
  h_{12}(\mathbf{ x})&=\frac{1-3(\mathbf n_{12}\cdot \mathbf n_1)^2}{2|\mathbf{ x}-\mathbf y_1|^3\epsilon}\nonumber\\
  &+\frac{2-9(\mathbf n_{12}\cdot \mathbf n_1)+15(\mathbf n_{12}\cdot \mathbf n_1)^3}{4|\mathbf{ x}-\mathbf y_1|^4}+\mathcal O(\epsilon)\ .\label{h12-h11}
\end{align}
We can finally average out $\mathbf n_{12}$ over spatial directions using $\langle n_{12}^i\rangle=0$, $\langle n_{12}^i n_{12}^j\rangle=\delta_{ij}/3$, and $\langle n_{12}^i n_{12}^j n_{12}^k\rangle=0$:
\begin{equation}
h_{11}(\mathbf{ x})=\frac{1}{2|\mathbf{ x}-\mathbf y_1|^4}\ .\label{h11}
\end{equation}

The simplicity of Eq.~(\ref{h11}) is striking: the Gauss-Bonnet contributions to the fields are {\em finite}, and no regularization procedure (see e.g.~\cite{Blanchet:2000cw}) is necessary to solve Eq.~(\ref{fieldEqn1PN}) at 1PN order. The generalization of this remarkable fact to higher PN orders is left to future work.

We can now solve Eq.~(\ref{fieldEqn1PN}) to find:
\begin{widetext}
\begin{subequations}
\begin{align}
U(x)&=\sum_A\frac{m_A^0}{\rho_A}\bigg[1+\frac{3}{2}\mathbf v_A^2-\sum_{B\neq A}(1+\alpha_A^0\alpha_B^0)\bigg]-4\alpha f'(\varphi_0)\sum_{A,B}m_A^0m_B^0\alpha_A^0 h_{AB}(\mathbf{ x})+\mathcal O(v^6)\ ,\\
g_i(x)&=\sum_A\frac{m_A^0 v_A^i}{|\mathbf{ x}-\mathbf x_A(t)|}+\mathcal O(v^5)\ ,\\
\varphi(x)&=\varphi_0-\sum_A\frac{m_A^0\alpha_A^0}{\rho_A}\bigg[1-\frac{1}{2}\mathbf v_A^2-\sum_{B\neq A}\frac{m_A^0}{r}\left(1+\alpha_A^0\alpha_B^0-\frac{\beta_A^0\alpha_B^0}{\alpha_A^0}\right)\bigg]+2\alpha f'(\varphi_0)\sum_{A,B}m_A^0m_B^0\, h_{AB}(\mathbf{ x})+\mathcal O(v^6)\ ,
\end{align}\label{FinalFieldsPN}
\end{subequations}
\end{widetext}
where $x^\mu=(t,\mathbf{ x})$ and
\begin{align}
\frac{1}{\rho_A}&=\frac{1}{|\mathbf{ x}-\mathbf x_A(t)|}+\frac{1}{2}\partial_t^2|\mathbf{ x}-\mathbf x_A(t)|\nonumber\\
                &=\frac{1}{|\mathbf{ x}-\mathbf x_A(t)|}\left[1+\frac{1}{2}\mathbf v_A^2-\frac{1}{2}(\mathbf n_A\cdot \mathbf v_A)^2\right]\nonumber\\
                  &+\frac{1}{2}(\mathbf n_A\cdot \mathbf a_A)\ ,\label{DefRhoPN}
\end{align}
with $\mathbf n_A=(\mathbf x_A-\mathbf{ x})/|\mathbf x_A-\mathbf{ x}|$ and $\mathbf a_A=d\mathbf v_A/dt$.

The two-body Lagrangian can now be straightforwardly obtained \`a la Droste-Fichtenholz, a technique which, at this order, is equivalent to computing, e.g., a Fokker Lagrangian \cite{Blanchet:2013haa}. First, one writes the Lagrangian of, say, body $A$ considered as a test particle in the fields of $B$:
\begin{align}
  L_A&=-m_A(\varphi)\frac{ds_A}{dt}\label{LAdef}\\
  &=-m_A(\varphi)\sqrt{e^{-2U}+8g_i v^i_A-e^{2U}\mathbf v_A^2}+\mathcal O(v^6)\ , \nonumber
\end{align}
where $U$, $g_i$ and $\varphi$ are given by (\ref{FinalFieldsPN}), setting formally $m_A^0=0$ and $\mathbf{ x}=\mathbf x_A$. In particular, Eq.~(\ref{DefRhoPN}) can be rewritten as
\begin{align}
\frac{1}{\rho_A}&=\frac{1}{R}\left[1+\frac{1}{2}(\mathbf v_A\cdot \mathbf v_B)-\frac{1}{2}(\mathbf n\cdot \mathbf v_A)(\mathbf n\cdot \mathbf v_B)\right]\nonumber\\
&+\frac{1}{2}\frac{d}{dt}(\mathbf n\cdot \mathbf v_A)\ ,
\end{align}
with $r=|\mathbf x_A-\mathbf x_B|$ and $\mathbf n=(\mathbf x_A-\mathbf x_B)/r$. Note that the last term is a total time derivative, that can be ignored in the Lagrangian (\ref{LAdef}).

The final two-body Lagrangian $L_{AB}$ is easily inferred from $L_A$. Indeed, the only Lagrangian that is symmetric under exchange of the bodies ($A\leftrightarrow B$), and whose resulting equations of motion reduce to those of $L_A$ in the test-mass limit $m_A^0\ll m_B^0$ is:
\begin{widetext}
\begin{align}
L_{AB}&=-m_A^0-m_B^0+\frac{1}{2}m_A^0 \mathbf v_A^2+\frac{1}{2}m_B^0 \mathbf v_B^2+\frac{m_A^0m_B^0}{r}(1+\alpha_A^0\alpha_B^0)+\frac{1}{8}m_A^0 \mathbf v_A^4+\frac{1}{8}m_B^0 \mathbf v_B^4 \nonumber \\
&+\frac{m_A^0m_B^0}{r}\left[\left(\frac{\mathbf v_A\cdot \mathbf v_B}{2}(-7+\alpha_A^0\alpha_B^0)\right)+\left(\frac{ \mathbf v_A^2+\mathbf v_B^2}{2}(3-\alpha_A^0\alpha_B^0)\right)-\left(\frac{(\mathbf n\cdot \mathbf v_A)(\mathbf n\cdot \mathbf v_B)}{2}(1+\alpha_A^0\alpha_B^0)\right)\right] \nonumber \\
&-\frac{m_A^0m_B^0}{2r^2}\left[m_A^0\left((1+\alpha_A^0\alpha_B^0)^2+\beta_B^0{\alpha_A^0}^2\right)+m_B^0\left((1+\alpha_A^0\alpha_B^0)^2+\beta_A^0{\alpha_B^0}^2\right)\right] \nonumber \\
&+\frac{\alpha f'(\varphi_0)}{r^2}\frac{m_A^0m_B^0}{r^2}\left[m_A^0(\alpha_B^0+2\alpha_A^0)+m_B^0(\alpha_A^0+2\alpha_B^0)\right]+\mathcal O(v^6)\ ,
\end{align}
\end{widetext}
which is straightforwardly rewritten as Eq.~(\ref{1PNLagrangian}).

This completes our derivation.

\begin{figure*}[th]
  \includegraphics[width=\columnwidth]{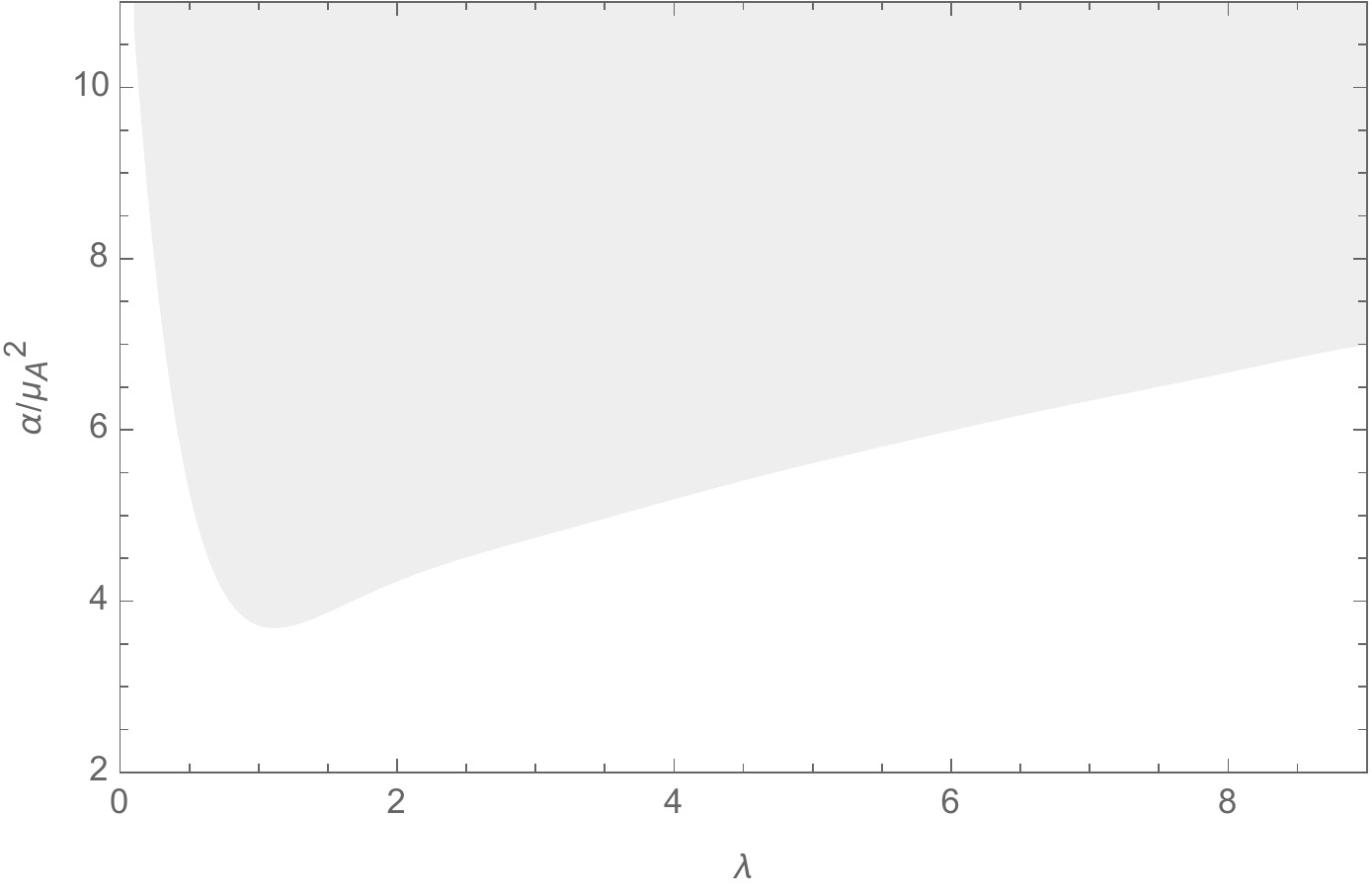}
  \includegraphics[width=\columnwidth]{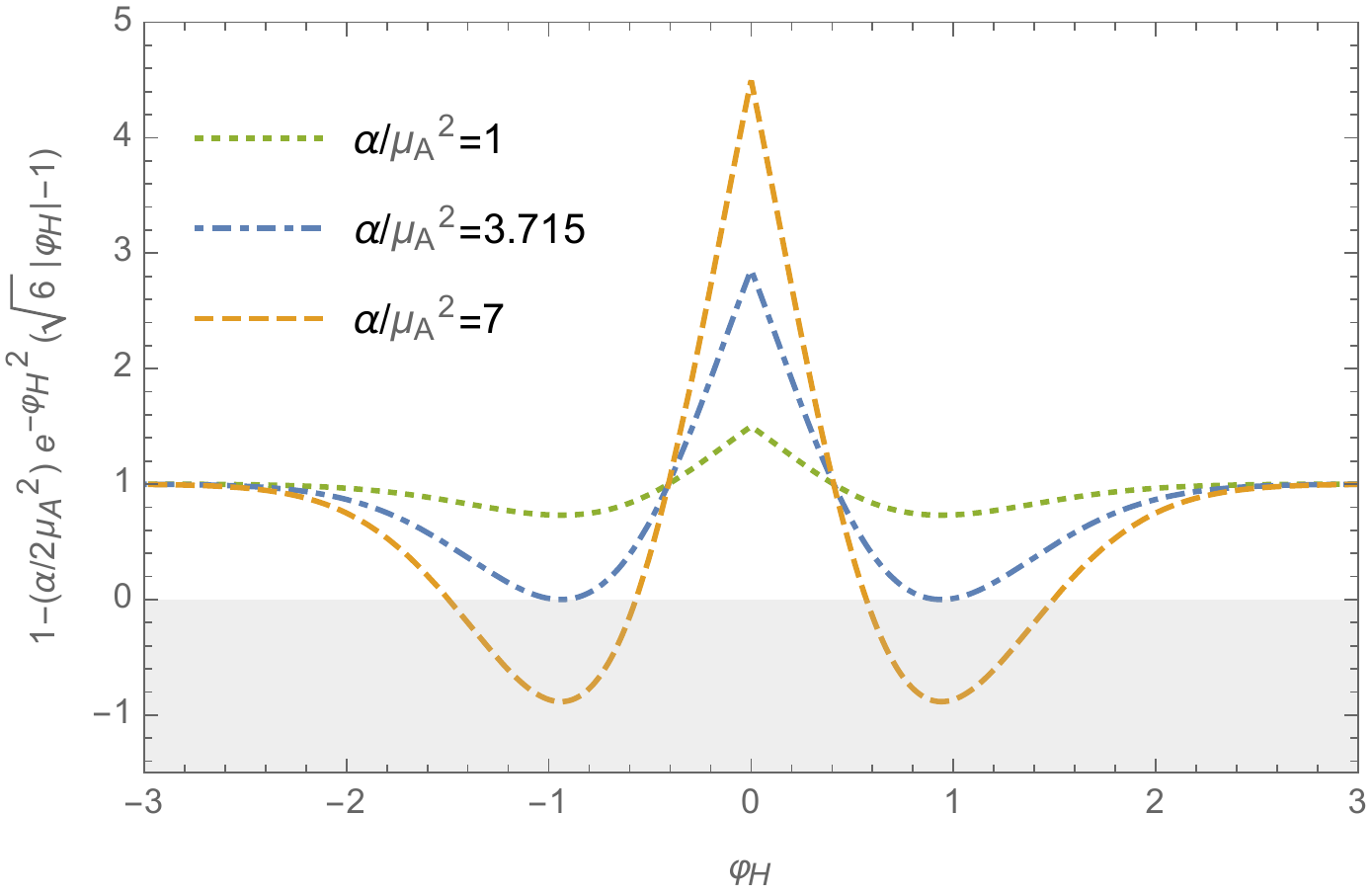}
\caption{Left panel: Parameter space $\{\alpha/\mu_A^2,\lambda\}$ in quadratic EsGB theory. The white area represents the parameter space for which the bound (\ref{BHconditionQuadratic}) is satisfied $\forall\,\varphi_{\rm H}$. At the boundary with the shaded area, (\ref{BHconditionQuadratic}) has two symmetric roots in the $\varphi_{\rm H}$ variable. In the shaded area, (\ref{BHconditionQuadratic}) is violated within two symmetric $\varphi_{\rm H}$ intervals. Right panel: The example $\lambda=1$. The bound (\ref{BHconditionQuadratic}) is violated in two symmetric $\varphi_{\rm H}$ intervals when $\alpha/\mu_A^2>3.715$. 
\label{quad_criticalRatio}}
\end{figure*}

\begin{figure*}[th]
\centering
  \includegraphics[width=\columnwidth]{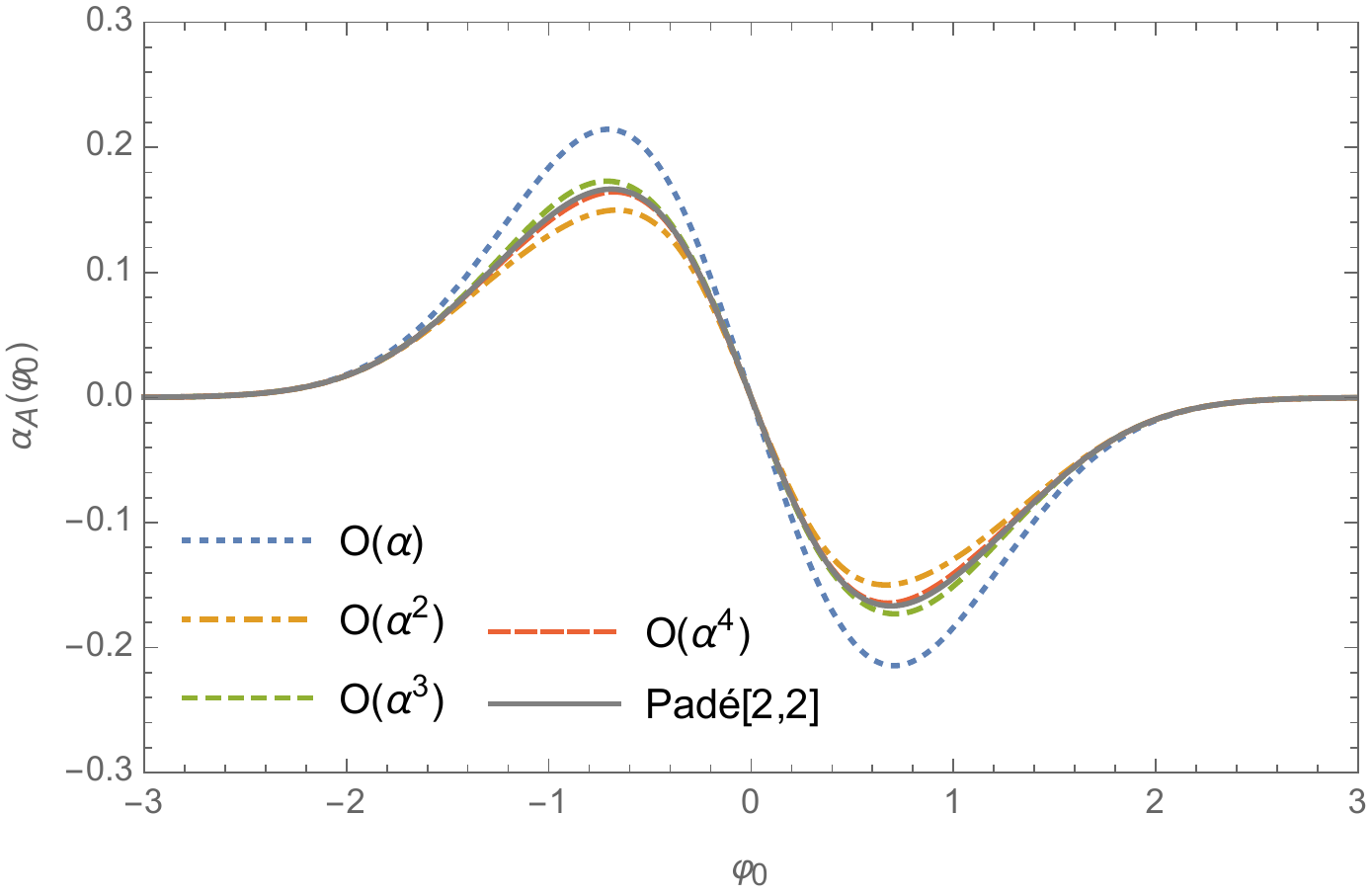}
  \includegraphics[width=\columnwidth]{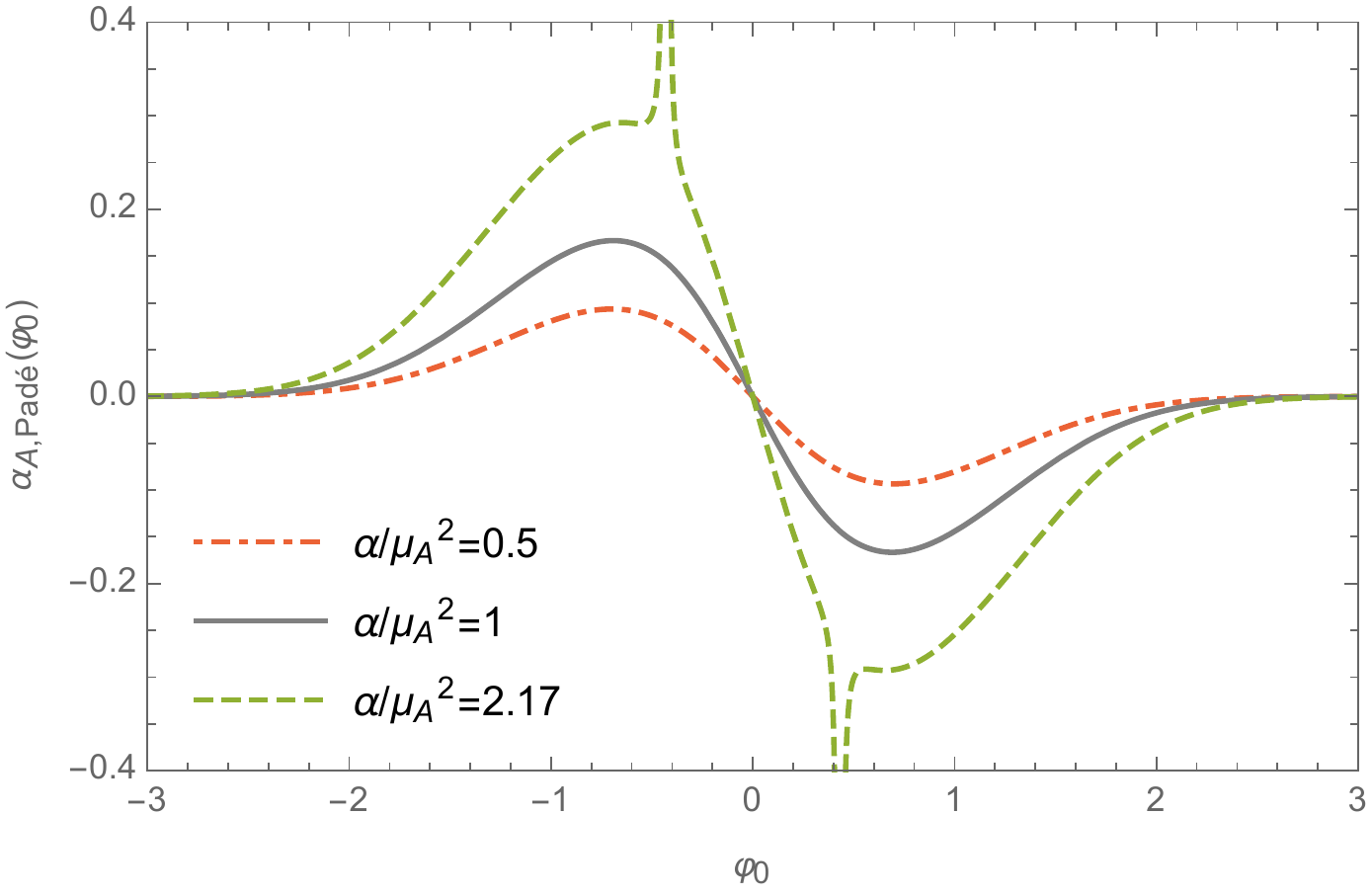}
\caption{Scalar coupling $\alpha_A(\varphi_0)$ of BHs in theories with quadratic coupling of the form (\ref{couplingQuadratic}) with $\lambda=1$. Left panel: Taylor series (\ref{scalarCouplingQuadratic}) truncated at order $\mathcal O(x^n)$ and its $(2,2)$ Pad\'e resummation $\alpha_{A,\text{Pad\'e}}^0$, for the special case $\alpha/\mu_A^2=1$. Right panel: $\alpha_{A,\text{Pad\'e}}(\varphi_0)$ for three different BHs with $\alpha/\mu_A^2=\{0.5, 1, 2.17\}$. When $\alpha/\mu_A^2$ becomes larger than the critical value corresponding to $(\alpha/\mu_A^2)_{\rm crit}^{\text{Pad\'e}}=2.17$, two singularities appear at $\varphi_0^{\rm crit}=\pm 0.42$.
\label{quadratic_figure}}
\end{figure*}

\section{Sensitivities for quadratic coupling and shift-symmetric theories}
\label{sec:AppQuadraticShiftSymmetric}

In Sec.~\ref{PNparameters} we studied BH sensitivities in one of the best motivated subclasses of EsGB theory, namely EdGB gravity. Here we generalize the analysis to quadratic and shift-symmetric EsGB theories.

\subsubsection{Quadratic coupling\label{subsectionQuadratic}}
Let us consider EsGB theories where the coupling function depends only on $\varphi^2$, i.e. is of the form~\cite{Doneva:2017bvd}
\begin{equation}
  f(\varphi)=-\frac{e^{-\lambda\varphi^2}}{2\lambda}
  \label{couplingQuadratic}
\end{equation}
with $\lambda>0$. The EsGB action (\ref{vacuumAction}) is symmetric under $\varphi\to -\varphi$.
The coefficients appearing in the scalar coupling function $\alpha_A^0$ [cf. Eqs.~(\ref{scalarChargeUnexpanded}) and (\ref{scalarCouplingA})] now read
\begin{widetext}
\begin{subequations}
\begin{align}
A_2^{\rm quad}(\varphi_0)&=\frac{-120+73\lambda}{480\lambda\varphi_0}-\frac{73\lambda\varphi_0}{240}\ ,\\
A_3^{\rm quad}(\varphi_0)&=\frac{110376 \lambda^2 - 87577 \lambda^3}{
 241920 \lambda^2} + \frac{30240 - 36792 \lambda + 12511 \lambda^2}{241920 \lambda^2 \varphi_0^2} + \frac{
 12511 \lambda^2 \varphi_0^2}{40320}\ ,\\
A_4^{\rm quad}(\varphi_0)&=\frac{-798336000 + 1456963200 \lambda - 990871200 \lambda^2 + 227192473 \lambda^3}{
 12773376000 \lambda^3 \varphi_0^3}\\
 & + \frac{
 -5827852800 \lambda^2 + 9578872320 \lambda^3 - 3685838076 \lambda^4}{
 12773376000 \lambda^3 \varphi_0} +\frac{(-11230775040 \lambda^4 + 9239974444 \lambda^5) \varphi_0}{
 12773376000 \lambda^3} -\frac{ 102384391 \lambda^3 \varphi_0^3}{266112000}\ .\nonumber
\end{align}
\end{subequations}
\end{widetext}

In the special case $\lambda=1$ we find
\begin{align}
  \alpha_A^0&=-\frac{x}{2}+\left(\frac{47}{480 \varphi_0 }+\frac{73 \varphi_0 }{240}\right)
   x^2\label{scalarCouplingQuadratic}\\
   &+\left(\frac{3257}{34560}+\frac{5959}{241920 \varphi_0^2}+\frac{12511 \varphi_0
   ^2}{40320}\right) x^3\nonumber\\
&+\left(\frac{15007361}{1824768000 \varphi_0
   ^3}-\frac{5431787}{1064448000 \varphi_0 }+\frac{497700149 \varphi_0
   }{3193344000}\right.\nonumber\\
   &\hspace*{3cm}\left.+\frac{102384391 \varphi_0 ^3}{266112000}\right) x^4+\mathcal O(x^5)\nonumber
\end{align}
with
\begin{equation}
  x=\frac{ \alpha (\varphi_0\,  e^{-\varphi_0^2}) }{\mu_A^2}\ .\nonumber
\end{equation}
As expected, under a sign inversion $\varphi_0\to -\varphi_0$ we have $\alpha_A^0\to -\alpha_A^0$ and $\beta_A^0=(d\alpha_A/d\varphi)(\varphi_0)\to \beta_A^0$, so that the two-body Lagrangian [Eqs.~(\ref{1PNLagrangian}) and (\ref{defCombinaisonsPN})] is invariant.

For EsGB theories with quadratic couplings of the form (\ref{couplingQuadratic}), a BH with irreducible mass $\mu_A$ is regular outside the horizon if the condition (\ref{BHbound}) is satisfied, i.e. if
\begin{equation}
\frac{\alpha e^{- \lambda\varphi_{\rm H}^2}}{2\mu_A^2}\left(\sqrt{6}|\varphi_{\rm H}|-\frac{1}{\lambda}\right)<1\ .\label{BHconditionQuadratic}
\end{equation}
For $\lambda=1$, this condition is satisfied for all $\varphi_{\rm H}$ whenever $\alpha/\mu_A^2<(\alpha/\mu_A^2)_{\rm crit}= 3.715$, and then BH $A$ can never reach the singular configuration, whatever the value of the background scalar field $\varphi_0$.

Note that the condition above is not very restrictive, as the coupling constant $\alpha/\mu_A^2$ is expected to be small. The same conclusions apply to the case $\lambda\neq 1$, as illustrated in the left panel of Fig.~\ref{quad_criticalRatio}. In the white region of the $\{\alpha/\mu_A^2,\lambda\}$ plane, the inequality (\ref{BHconditionQuadratic}) is satisfied for any value $\varphi_{\rm H}$ of the scalar field at the horizon. In the shaded area, the inequality (\ref{BHconditionQuadratic}) is violated within two symmetric $\varphi_{\rm H}$ intervals. At the boundary between these two regions, these intervals reduce to two points.

The right panel focuses on the special case $\lambda=1$. When $\alpha/\mu_A^2>(\alpha/\mu_A^2)_{\rm crit}$, the inequality (\ref{BHconditionQuadratic}) is violated when $\varphi_{\rm H}$ takes values in two intervals which are symmetric with respect to the origin. In the limit $\alpha/\mu_A^2\to +\infty$, these intervals tend to $]-\infty,-\frac{1}{\lambda\sqrt{6}}]$ and $[\frac{1}{\lambda\sqrt{6}},+\infty[$, respectively.

Figure~\ref{quadratic_figure}, which is completely analogous to Fig.~\ref{EdGB_scalarCoupling}, shows $\alpha_A^0(\varphi_0)$ for $\lambda=1$.  The left panel (where we set $\alpha/\mu_A^2=1$ for simplicity) shows that the Taylor series converges much faster than in the dilatonic case and that, unlike the dilatonic case, the sensitivity (\ref{scalarCouplingQuadratic}) is finite $\forall\varphi_0$.
The right panel shows the Pad\'e-resummed coupling $\alpha_A^{\text{Pad\'e}}$ when $\lambda=1$. The Pad\'e approximation suggests that two poles in $\alpha_A^{\text{Pad\'e}}(\varphi_0)$ appear at some critical coupling $(\alpha/\mu_A^2)_{\rm crit}^{\text{Pad\'e}}$ such that $(\alpha/\mu_A^2)_{\rm crit}^{\text{Pad\'e}}=2.17$. This value is qualitatively comparable to the nonperturbative prediction given below Eq.~(\ref{BHconditionQuadratic}). A more accurate estimate of $(\alpha/\mu_A^2)_{\rm crit}^{\text{Pad\'e}}$ using higher-order expansions in $\alpha$ is an interesting topic for future work.

\begin{figure*}[th]
  \includegraphics[width=\columnwidth]{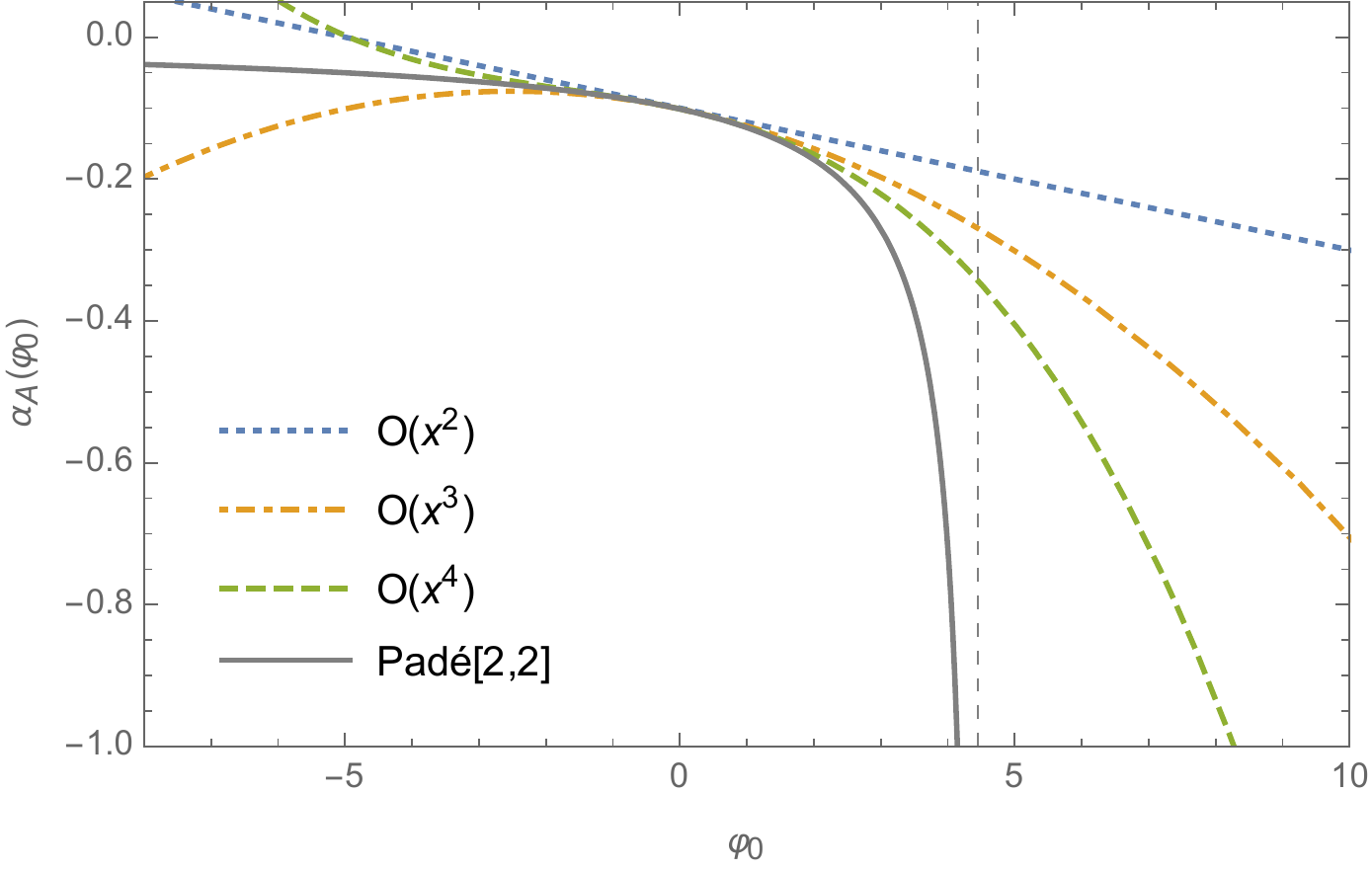}
  \includegraphics[width=\columnwidth]{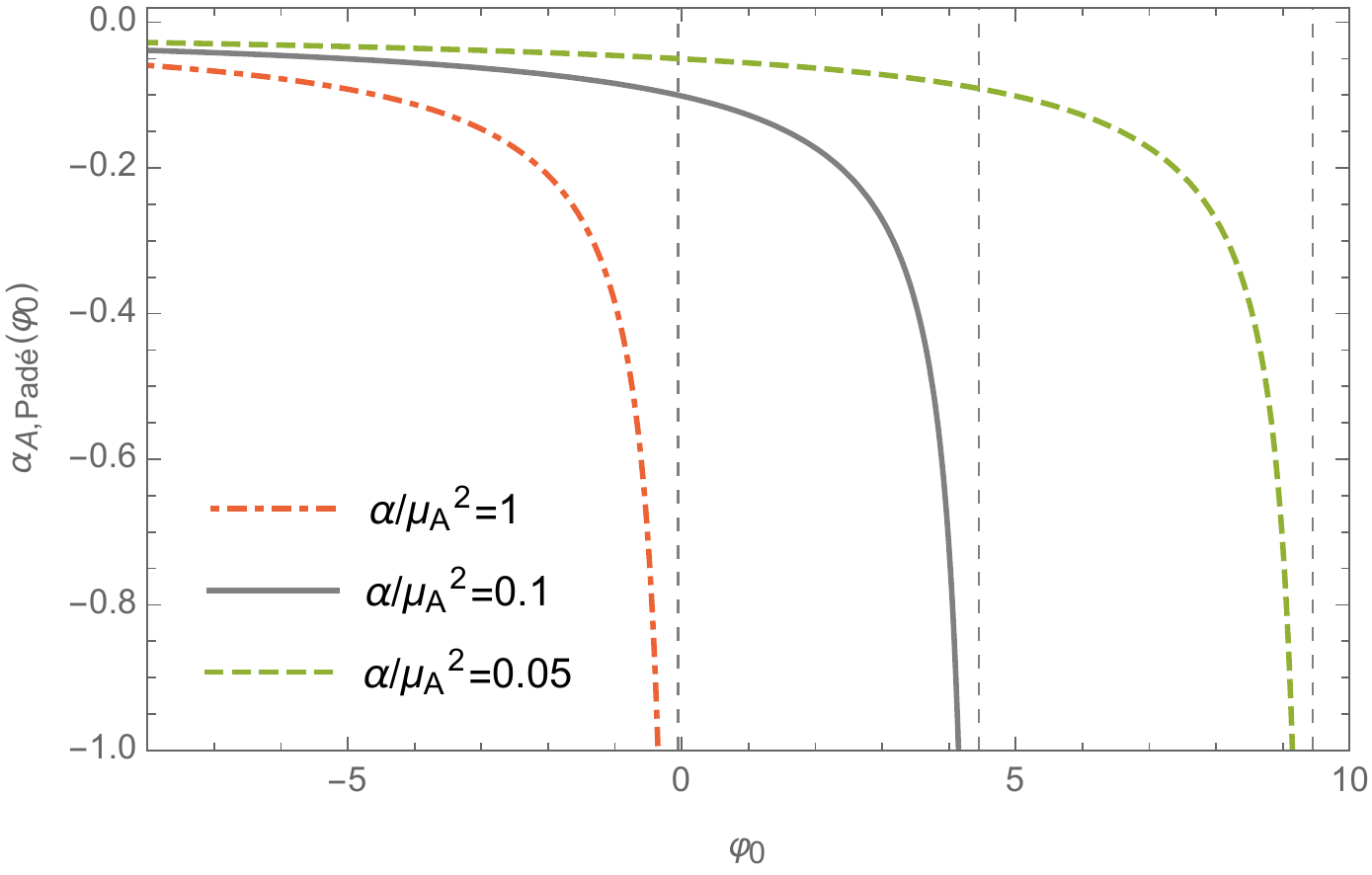}
  \label{fig:Quad2}
\caption{Scalar coupling $\alpha_A(\varphi_0)$ of BHs in the shift-symmetric theory (\ref{couplingShift}). Left panel: Taylor series (\ref{scalarCouplingShiftSym}) truncated at order $\mathcal O(x^n)$ and its $(2,2)$ Pad\'e resummation $\alpha_{A,\text{Pad\'e}}^0$, for the special case $\alpha/\mu_A^2=0.1$. Right panel: $\alpha_{A,\text{Pad\'e}}^0$ for three different BHs with $\alpha/\mu_A^2=\{1, 0.1, 0.05\}$.
\label{shiftSym_figure}}
\end{figure*}

Once again, the scalar field value $\varphi_0$ at infinity plays a major role. As $|\varphi_0|$ increases, the sensitivity $|\alpha_A^0|$ also does, until it reaches an extremum at
\begin{equation}
  \varphi_0^{\rm extr}=\pm\frac{1}{\sqrt{2\lambda}}+\mathcal O(x)\,,
\end{equation}
where
\begin{align}
  \alpha_A(\varphi_0^{\rm extr})&=\mp\frac{\alpha}{2\mu_A^2}\frac{1}{\sqrt{2e\lambda}}+\mathcal O(x^2)
\end{align}
and $\beta_A(\varphi_0^{\rm ext})=0$. Here $e$ is Euler's number. In the limit $|\varphi_0|\gg|\varphi_0^{extr}|$, instead, $\alpha_A^0\to 0$ and $\beta_A^0=(d\alpha_A/d\varphi)(\varphi_0)\to 0$, so the BH is indistinguishable from the Schwarzschild solution.
Finally, the sensitivity ``turns off'' when $\varphi_0=0$: $\alpha_A(0)=0$. This is because $\alpha_A^0$ is associated to the BH solutions of Sec.~\ref{sectionTNsolution}, which were derived in the weak Gauss-Bonnet coupling limit: see Eq.~(\ref{SSSansatz}). When $\varphi_0=0$, $f'(\varphi_\infty)=0$ and the solution reduces to the Schwarzschild metric. Note that the branch of ``spontaneously scalarized'' BH solutions with nonperturbative scalar hair and $\varphi_\infty=0$~\cite{Silva:2017uqg,Doneva:2017bvd,Cunha:2019dwb} is not included in our analysis. A numerical calculation of their sensitivities $\alpha_A^0$ and $\beta_A^0$ is left for future work.

\subsubsection{Shift-symmetric theories\label{subsectionShiftSym}}

As a third and last example, let us consider shift-symmetric theories~\cite{Sotiriou:2013qea,Sotiriou:2014pfa} with
\begin{equation}
f(\varphi)=2\varphi\ .\label{couplingShift}
\end{equation}
The action (\ref{vacuumAction}) is symmetric under the shift symmetry $\varphi\to \varphi+\Delta\varphi$, where $\Delta\varphi$ is a constant. The sensitivity (\ref{scalarChargeUnexpanded}) reads
\begin{align}
\alpha_A^0&=-\frac{x}{2}-\frac{\varphi_0}{2}x^2-\left(\frac{73}{480}+\frac{{\varphi_0}^2}{2}\right)x^3 \nonumber \\
          &-\left(\frac{73\varphi_0}{160}+\frac{{\varphi_0}^3}{2}\right)x^4+\mathcal O(x^5)\label{scalarCouplingShiftSym}
\end{align}
with
\begin{equation}
            x=\frac{2 \alpha }{\mu_A^2}\,,
\end{equation}
and it is also invariant under $\varphi_0\to\varphi_0+\Delta\varphi$, since then $\mu_A^2=S_{\rm w}/4\pi\to \mu_A^2+2\alpha\Delta\varphi$: cf. Eq.~(\ref{waldEntropy2}).

In Fig.~\ref{shiftSym_figure} we plot $\alpha_A^0$ as a function of $\varphi_0$. The left panel (where we set $\alpha/\mu_A^2=0.1$ for concreteness) shows that the series (\ref{scalarCouplingShiftSym}) converges on a narrow interval. When $\varphi_0$ is large and positive, $\alpha_A^0$ diverges with a slope which increases with the truncation order $\mathcal O(x^n)$; when $\varphi_0$ is large and negative, $\alpha_A^0$ diverges, but $\text{sign}(\alpha_A^0)=(-1)^n$ depends on the truncation order.  To improve the convergence properties of the expansion (\ref{scalarCouplingShiftSym}), we try a diagonal $(2,2)$ Pad\'e resummation, also shown in the left panel of Fig.~\ref{shiftSym_figure}. The features of the Pad\'e resummation resemble the dilatonic case of Sec.~\ref{PNparameters}:
\begin{itemize}
  \item[(i)] when $\varphi_0\to -\infty$ the BH decouples from the scalar field, i.e. $\alpha_A^0\to 0$ and $\beta_A^0=(d\alpha_A^0/d\varphi)(\varphi_0)\to 0$;
  \item [(ii)] as $\varphi_0$ increases , the BH becomes strongly coupled to the scalar field: $\alpha_A^0\to -\infty$ and $\beta_A^0\to -\infty$ as $\varphi_0$ approaches a pole located at
\begin{equation}
\varphi_0^{\rm pole}=\frac{1}{2}\left(\frac{\mu_A^2}{\alpha}-\frac{\sqrt{1095}}{30}\right)\ .\label{poleLocationShiftSym}
\end{equation} 
\end{itemize}
Once again, $\varphi_0$ plays a crucial role. The BH's irreducible mass $\mu_A$ only affects the location of the pole through Eq.~(\ref{poleLocationShiftSym}), as shown in the right panel of Fig.~\ref{shiftSym_figure}.
The features highlighted above are again valid within the nonperturbative bound (\ref{BHbound}), which now reads
\begin{equation}
\varphi_{\rm H}<\frac{1}{2}\left(\frac{\mu_A^2}{\alpha}-\sqrt{6}\right)\ .
\end{equation}
This equation predicts the existence of a maximum value for $\varphi_{\rm H}$ which depends linearly on $\alpha/\mu_A^2$.
A numerical study and higher-order expansions in $\alpha$, possibly combined with Pad\'e resummation techniques, would be useful to confirm these predictions.

\bibliography{edgbib}

\end{document}